\documentclass[iop,revtex4]{emulateapj}
\usepackage{natbib}
\usepackage{graphicx}
\usepackage{float}
\usepackage{lscape}
\usepackage{color}
\shorttitle{Orbits for the Impatient}
\shortauthors{Blunt, Nielsen, De Rosa, et al.}

\begin{document}
\submitted{Accepted to AJ on 03/23/2017}
\title{Orbits for the Impatient: A Bayesian Rejection Sampling Method for Quickly Fitting the Orbits of Long-Period Exoplanets}

\author{
Sarah Blunt\altaffilmark{1,2,3},
Eric L. Nielsen\altaffilmark{2,3},
Robert J. De Rosa\altaffilmark{4},
Quinn M. Konopacky\altaffilmark{5},
Dominic Ryan\altaffilmark{4},
Jason J. Wang\altaffilmark{4},
Laurent Pueyo\altaffilmark{6},
Julien Rameau\altaffilmark{7}, 
Christian Marois\altaffilmark{8},
Franck Marchis\altaffilmark{2},
Bruce Macintosh\altaffilmark{3},
James R. Graham\altaffilmark{4},
Gaspard Duch{\^e}ne\altaffilmark{4,9},
and Adam C. Schneider\altaffilmark{10}
}

\altaffiltext{1}{Department of Physics, Brown University, Providence, RI 02912, USA}

\altaffiltext{2}{SETI Institute, Carl Sagan Center, 189 Bernardo Avenue, Mountain View, CA 94043, USA}

\altaffiltext{3}{Kavli Institute for Particle Astrophysics and Cosmology, Stanford University, Stanford, CA 94305, USA}

\altaffiltext{4}{Astronomy Department, University of California, Berkeley, CA 94720, USA}

\altaffiltext{5}{Center for Astrophysics and Space Science, University of California San Diego, La Jolla, CA 92093, USA}

\altaffiltext{6}{Space Telescope Science Institute, Baltimore, MD 21218, USA}

\altaffiltext{7}{Universit{\'e} de Montr{\'e}al, Montr{\'e}al, QC, H3T 1J4, Canada}

\altaffiltext{8}{National Research Council of Canada, Victoria, V9E 2E7, Canada}

\altaffiltext{9}{Universit{\'e} Grenoble Alpes / CNRS, Institut de Plan{\'e}tologie d'Astrophysique de Grenoble, 38000 Grenoble, France}

\altaffiltext{10}{School of Earth and Space Exploration, Arizona State University, Tempe, AZ 85287, USA}

\begin{abstract}
We describe a Bayesian rejection sampling algorithm designed to efficiently compute posterior distributions of orbital elements for data covering short fractions of long-period exoplanet orbits. Our implementation of this method, Orbits for the Impatient (OFTI), converges up to several orders of magnitude faster than two implementations of MCMC in this regime. We illustrate the efficiency of our approach by showing that OFTI calculates accurate posteriors for all existing astrometry of the exoplanet 51 Eri b up to 100 times faster than a Metropolis-Hastings MCMC. We demonstrate the accuracy of OFTI by comparing our results for several orbiting systems with those of various MCMC implementations, finding the output posteriors to be identical within shot noise. We also describe how our algorithm was used to successfully predict the location of 51 Eri b six months in the future based on less than three months of astrometry. Finally, we apply OFTI to ten long-period exoplanets and brown dwarfs, all but one of which have been monitored over less than 3\,\% of their orbits, producing fits to their orbits from astrometric records in the literature. 
\end{abstract}


\section{Introduction}

Direct imaging is sensitive to substellar objects with large projected separations from their host objects ($\gtrsim$0.2\,"; e.g. \citealt{Macintosh:2014}, \citealt{Claudi:2016}), corresponding to larger orbital semi-major axes and periods compared to those detected with radial velocity and transit methods \citep{Bowler:2016}. Therefore, over timescales of months to years, direct imaging observations often probe only short fractions of these orbits. In these cases, constraints on orbital parameters can be used to perform a preliminary characterization of the orbit (e.g. \citealt{Beust:2014}, \citealt{Nielsen:2014}, \citealt{Millar-Blanchaer:2015}, \citealt{Sallum:2015}, \citealt{De Rosa:2015}, \citealt{Zurlo:2016}, \citealt{Rameau:2016}). Orbital parameter constraints can also lead to mass constraints on directly imaged substellar objects (e.g. \citealt{Lagrange:2012}, \citealt{Fabrycky:2010}), constraints on additional planets in the system \citep{Bryan:2016}, and information about the interactions between planets and circumstellar disks (e.g. \citealt{Nielsen:2014}, \citealt{Millar-Blanchaer:2015}, \citealt{Rameau:2016}). In addition, orbit fitting can be used to constrain the future locations of exoplanets, notably to calculate the probability of a transit (e.g. \citealt{Wang:2016}), or to determine an optimal cadence of observations to reduce uncertainty in orbital parameter distributions. For future direct imaging space missions such as the \textit{Wide-Field Infrared Survey Telescope} (\textit{WFIRST}; \citealt{Spergel:2015}, \citealt{Traub:2016}), it is particularly important to quickly and accurately fit newly discovered exoplanet orbits in order to plan future observations efficiently.  

Several orbital fitting methods are currently used in astronomy. The family of Bayesian Markov Chain Monte Carlo methods (MCMC) was introduced to the field of exoplanet orbit fitting by Ford (2004, 2006) and has been widely used (e.g. \citealt{Nielsen:2014}, \citealt{Millar-Blanchaer:2015}, \citealt{Dupuy:2016}). MCMC is designed to quickly locate and explore the most probable areas of parameter space for a particular set of data, and takes longer to converge as a parameter space becomes less constrained by data, as in the case of astrometry from a fraction of a long-period orbit. In addition, many types of MCMC algorithms can be inefficient at exploring parameter spaces if the corresponding $\chi^2$ surface is complicated (e.g. \citealt{Ford:2004}). Another commonly used tool for fitting orbits is the family of least-squares Monte Carlo (LSMC) methods \citep{Press:1992}, which uses a Levenberg-Marquardt minimization algorithm to locate the orbital fit with minimum $\chi^2$ value for a set of astrometry. Once the minimum $\chi^2$ orbit is discovered, this method then randomly varies the measured astrometry along Gaussian distributions defined by the observational errors. In cases where the parameter space is very unconstrained, this method often explores only the area closest to the minimum $\chi^2$ orbit, leading to biases against  areas of parameter space with lower likelihoods. For example, \citet{Chauvin:2012} found significantly different families of solutions when using LSMC than when using MCMC for the same orbital data for $\beta$ Pic b. LSMC is therefore effective at finding the best-fit solution, but not well-suited to characterizing uncertainty by fully exploring the parameter space. 

In this work, we present Orbits for the Impatient (OFTI), a Bayesian Monte Carlo rejection sampling method based on that described in \citet{Ghez:2008}, and similar to the method described in \citet{Konopacky:2016}. OFTI is designed to quickly and accurately compute posterior probability distributions from astrometry covering a fraction of a long-period orbit. We describe how OFTI works and demonstrate its accuracy by comparing OFTI to two independent MCMC orbit-fitting methods. We then discuss situations where OFTI is most optimally used, and apply OFTI to several sets of astrometric measurements from the literature.

\section{The OFTI Algorithm}
\subsection{Method}

OFTI, like other Bayesian methods, combines astrometric observations and uncertainties with prior probability density functions (PDFs) to produce posterior PDFs of orbital parameters. These orbital parameter posteriors allow us to better characterize systems, for example by predicting future motion or by directly comparing the orbital plane to the orbits of other objects in the system or the distribution of circumstellar material.

The basic OFTI algorithm consists of the following steps:
\newline
\newline
1. Monte Carlo Orbit Generation from Priors
\newline
2. Scale-and-Rotate
\newline
3. Rejection sampling
\newline

OFTI uses a modified Bayesian rejection sampling algorithm. Rejection sampling consists of generating random sets of parameters, calculating a probability for each value, and preferentially rejecting values with lower probabilities. For OFTI, the generated parameters are the orbital elements semi-major axis, $a$, period, $P$, eccentricity, $e$, inclination angle, $i$, position angle of nodes, $\Omega$, argument of periastron, $\omega$, and epoch of periastron passage, $T_0$. For Bayesian rejection sampling algorithms such as OFTI, the candidate density functions used to generate these random parameters are prior probability distributions.

\subsubsection{Monte Carlo Orbit Generation from Priors} OFTI begins by generating an initial set of seven random orbital parameters drawn from prior probability distributions. In this work, we use a linearly descending eccentricity prior with a slope of -2.18 for exoplanets, derived from the observed distribution of exoplanets detected by the radial velocity method \citep{Nielsen:2010}. The use of this prior assumes that long-period exoplanets follow the same eccentricity distribution as the planets detected by the radial velocity method.While the shape of the eccentricity prior directly affects the shape of the eccentricity posterior, as we would expect, the posteriors of other parameters are less affected when changing between a linearly descending and a uniform prior (see section 3.1). We assume a purely random orientation of the orbital plane, which translates into a sin($i$) prior in inclination angle and uniform priors in the epoch of periastron passage and argument of periastron. That is, the inclination angle, position angle of nodes, and argument of periastron priors are purely geometric. OFTI initially generates orbits with $a$ = 1\,au  and $\Omega$ =  $0\,^{\circ}$, but these values are altered in the following step. We note that OFTI can be easily run using different priors, making it useful for non-planetary systems and statistical tests.

\subsubsection{Scale-and-Rotate} Once OFTI has generated an initial set of orbital parameters from the chosen priors, it performs a "scale-and-rotate" step in order to restrict the wide parameter space of all possible orbits. This increases the number of orbits accepted in the rejection sampling step. The generated semi-major axis and position angle of nodes are scaled and rotated, respectively, so that the new modified set of parameters describes an orbit that intersects a single astrometric data point. OFTI also takes the observational uncertainty of the data point used for the scale-and-rotate step into account. For each generated orbit, random offsets are introduced in separation ($\rho$) and position angle ($\theta$) from Gaussian distributions with standard deviations equal to the astrometric errors at the scale-and-rotate epoch. These offsets are added to the measured astrometric values, and then the generated orbit is scaled-and-rotated to intersect the offset data point, rather than the measured data point. The scale-and-rotate step produces a uniform prior in $\Omega$, and imposes a $\log(a)$ prior in semi-major axis. The posterior distributions OFTI produces are independent of the epoch chosen for this step, but the efficiency of the method is not. Some choices of the scale-and-rotate epoch result in a much higher fraction of considered orbits being accepted, and so the orbit is fit significantly faster. In order to take advantage of this change in efficiency, our implementation of OFTI performs an initial round of tests that pick out the scale-and-rotate epoch resulting in the largest acceptance rate of orbits, then uses this epoch every subsequent time this step is performed. The scale-and-rotate step differentiates OFTI from a true rejection sampling algorithm. 

\subsubsection{Rejection Sampling} Using the modified semi-major axis and position angle of nodes values, OFTI generates predicted $\rho$ and $\theta$ values for all remaining epochs. OFTI then calculates the $\chi^2$ probability for the predicted astrometry given the measured astrometry and uncertainties. This probability, assuming uncorrelated Gaussian errors, is given by: p  $\propto e^{-\chi^2 /2}$.

Finally, OFTI performs the rejection sampling step; it compares the generated probability to a number randomly chosen from a uniform distribution over the range (0,1). If the generated probability is greater than this random number, the generated set of orbital parameters is accepted.

\begin{figure}
\includegraphics[width=0.45\textwidth]{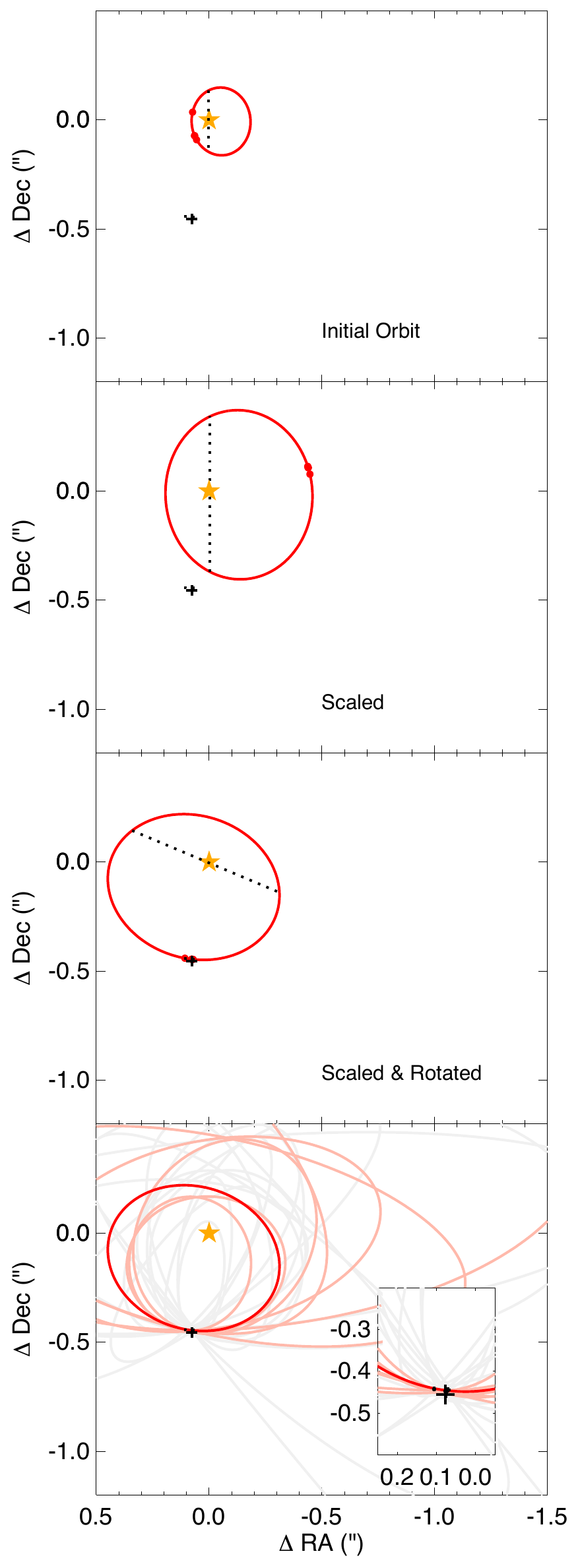}
\caption{Visualization of the OFTI method. Known astrometry (in this figure, for 51 Eri b), is shown as black crosses. The yellow star at (0,0) shows the location of 51 Eri A. Top: an orbit with $a$ = 1\,au, $\Omega$ =  $0\,^{\circ}$, and other orbital elements randomly drawn from appropriate priors is generated. Red dots along the orbit show the astrometric locations of 51 Eri b at the times of the observational epochs. The dotted line shows the line of nodes. Panels Two and Three: the scale-and-rotate step is performed. Bottom Panel: orbits accepted (light red) and rejected (gray) by the rejection sampling step of the OFTI algorithm. Inset: close up of bottom panel.}
\label{fig:oftiprocess}
\end{figure}

This process is repeated until a desired number of generated orbits has been accepted (see Figure \ref{fig:oftiprocess}). As with MCMC, histograms of the accepted orbital parameters correspond to posterior PDFs of the orbital elements.

Our implementation of OFTI makes use of several computational and statistical techniques to speed up the basic algorithm described above:

\begin{itemize}
\item 
Our implementation uses vectorized array operations rather than iterative loops wherever possible. For example, instead of generating one set of random orbital parameters at a time, our program generates arrays containing 10,000 sets of parameters, and performs all subsequent operations on these arrays. Our program then iterates over this main loop, accepting and rejecting in batches of 10,000 generated orbits at a time. 10,000 is the empirically determined optimal number for our implementation.
\item Our implementation of OFTI is written to run in parallel on multiple CPUs (our default is 10), speeding up runtime by a constant factor.
\item Our implementation of OFTI is equipped with a statistical speedup that increases the fraction of orbits accepted per orbits tested. Due to measurement errors, the minimum $\chi^2$ orbit typically has a non-reduced $\chi^2$ value greater than 0. OFTI makes use of this fact by calculating the minimum $\chi^2$ value of all orbits tested during an initial run, then subtracting this minimum value from all future generated $\chi^2$ values, rendering them more likely to be accepted in the method's rejection step. In rejection sampling, having a random variable whose range is greater than the maximum probability doesn't change the distribution of parameters, but does result in more rejected trials.

\item Our implementation of OFTI also restricts the ranges of the input $i$ and total mass priors based on initial results. After our implementation has accepted 100 orbits, it uses the maximum, minimum, and standard deviation of the array of accepted parameters to infer safe upper and lower limits to place on the relevant prior. This changes only the range of the relevant prior, not the shape of the prior. This speedup prevents our implementation of OFTI from wasting time generating orbits that have a negligible chance of being accepted. 

\end{itemize}

\subsection{Validation with MCMC}

\begin{figure}
\includegraphics[width=0.5\textwidth]{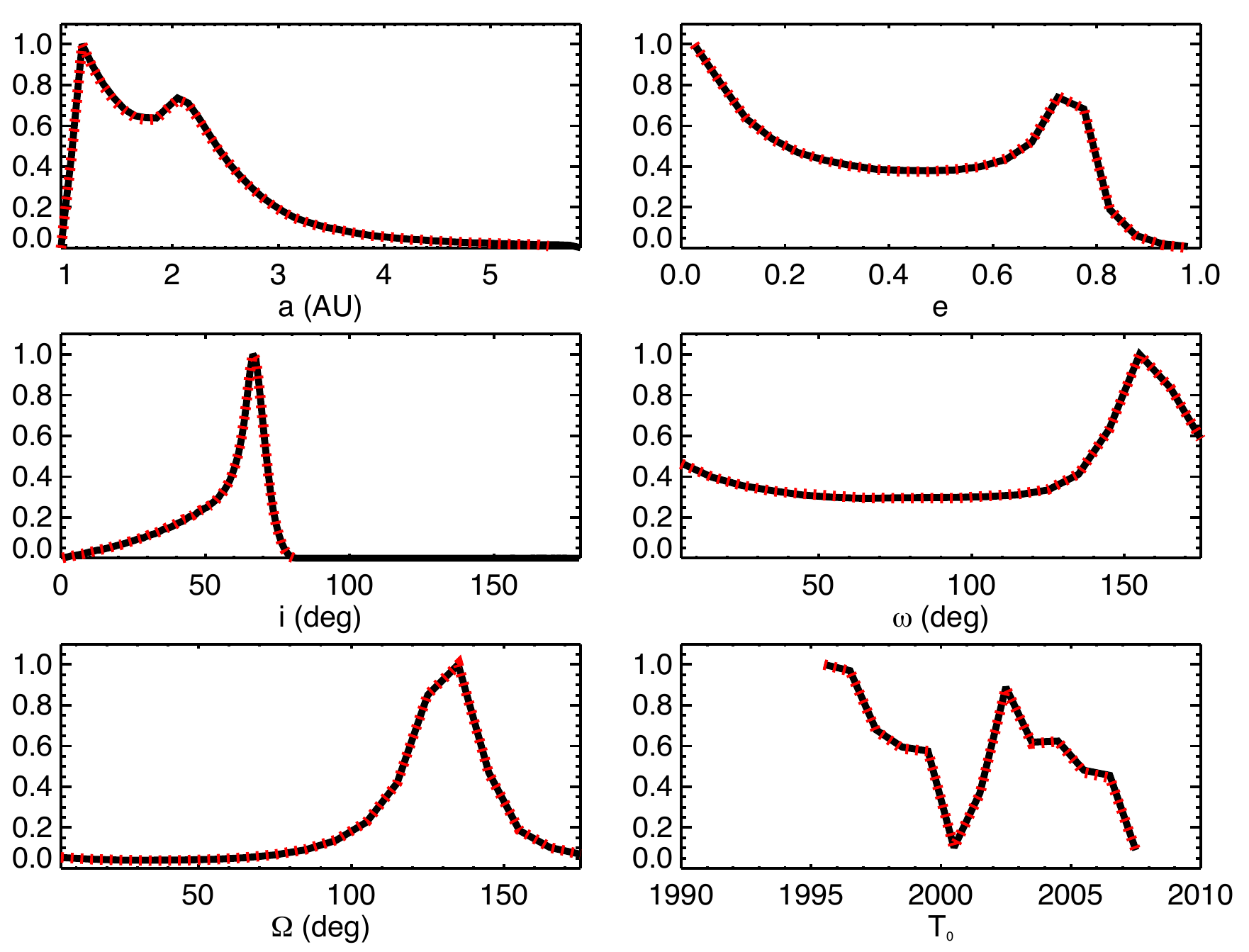}
\caption{Normalized PDFs of the orbital parameters corresponding to orbits accepted by OFTI (red) and MCMC (black) for astrometry of the ~9 year period binary brown dwarf SDSS 1052 from 2005-2006. The PDFs are identical, differing only by shot noise.}
\label{fig:sdss}
\end{figure}

\begin{figure}
\includegraphics[width=0.5\textwidth]{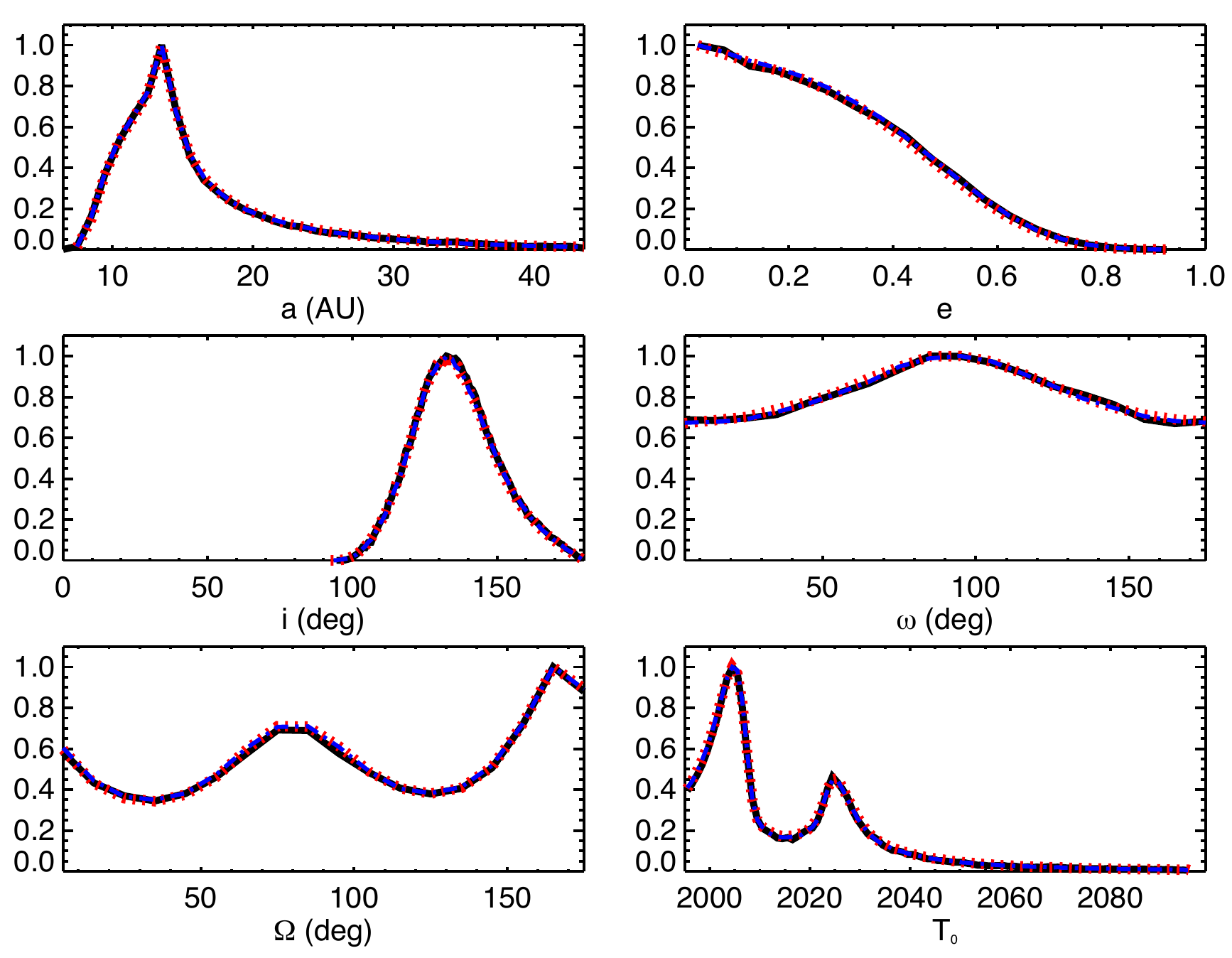}
\caption{Normalized PDFs of the orbital parameters corresponding to orbits accepted by OFTI (red), Metropolis-Hastings MCMC (black), and Affine-Invariant MCMC (blue) for relative astrometry of 51 Eri b from its discovery in 2014 December to 2015 September. The PDFs produced by all methods are identical, differing only by shot noise.}
\label{fig:goi2_compare}
\end{figure}

To illustrate that OFTI returns identical results to MCMC over short orbital arcs, we fit the same orbit and priors with OFTI and two MCMC orbit-fitting routines: the Metropolis Hastings MCMC algorithm described in \citet{Nielsen:2014}, and an Affine Invariant MCMC \citet{Foreman-Mackey:2013} orbit fitter from \citep{Macintosh:2014}. In Figure \ref{fig:sdss}, we plot the Metropolis Hastings MCMC and OFTI posterior PDFs calculated from astrometry of the system SDSS J105213.51+442255.7 AB (hereafter SDSS 1052; \citealt{Dupuy:2015}) from 2005-2006. SDSS 1052 is a pair of brown dwarfs with period of approximately 9 years. We chose only a subset of the available astrometry of SDSS 1052 to illustrate the effect of fitting a short orbital arc. In addition, we assume a fixed  system mass, and we use astrometry provided in Table 2 of \citealt{Dupuy:2015}. The posterior distributions produced by OFTI and the Metropolis-Hastings MCMC are identical. OFTI was also validated using the relative astrometry of 51 Eri b, a directly imaged exoplanet discovered by the GPI Exoplanet Survey in 2015 (\citealt{Macintosh:2015}, \citealt{De Rosa:2015}). In Figure \ref{fig:goi2_compare}, we plot the posterior distributions produced by all three methods, calculated from relative astrometry of 51 Eri b taken between 2014 December and 2015 September. As in the previous case, all three sets of posterior distributions produced by OFTI and the two MCMC implementations are identical.

\begin{figure}
\includegraphics[width=0.5\textwidth]{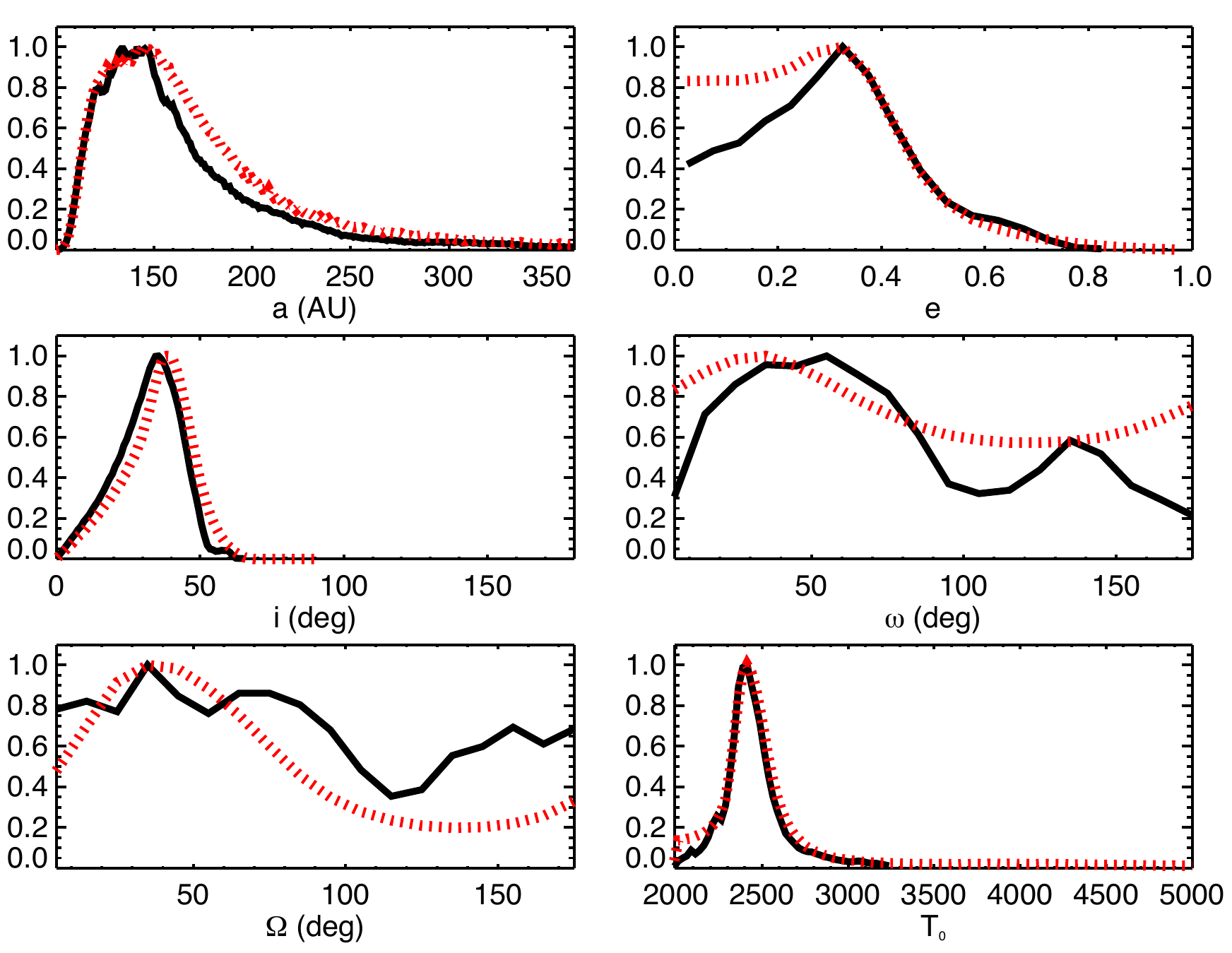}
\caption{Normalized PDFs of the orbital parameters produced by OFTI (red) and MCMC (black) for astrometry of ROXs 42B b. After approximately 30 hours of running in parallel on 10 cores, the MCMC chains are still unconverged, while OFTI produced $10^6$ permissible orbits in 134 minutes. The GR statistics for the MCMC chains plotted are all greater than 1.1, and the GR statistic for $\Omega$ is close to 1.5}
\label{fig:marta}
\end{figure}

An important difference between MCMC and OFTI involves the types of errors on the posteriors produced by the two methods. Because each step of OFTI is independent of previous steps, deviations from analytical posteriors have the form of uncorrelated noise, i.e. if our implementation of OFTI is run until $100$ orbits are accepted, the output posteriors will not be biased, but simply  noisy. As our implementation of OFTI is run until greater numbers of orbits are accepted, noise reduces by $\sqrt{N}$. MCMC steps, on the other hand, are not independent. Because the next MCMC step depends on the current location in parameter space, an un-converged MCMC run will result in biased, rather than noisy, posteriors. This is especially important in cases where MCMC has not been run long enough to achieve a satisfactory Gelman-Rubin (GR) statistic (a measure of convergence; \citealt{Gelman:1992}). In these cases, OFTI produces an unbiased result, while MCMC does not. This situation is illustrated in Figure \ref{fig:marta}, showing the posteriors produced by a Metropolis-Hastings implementation of MCMC and our implementation of OFTI for all known astrometry of ROXs 42B b (\citealt{Currie:2014}, \citealt{Bryan:2016}). After running for 30 hours on 10 CPUs, the MCMC posteriors are still un-converged, as can be seen by the lumpy shape of the $\Omega$ posterior. As a result, we see biases in the other posteriors. The GR statistics for each parameter were between 1.1 and 1.5 (an acceptable GR statistic is $\lesssim 1.01$, see e.g. \citealt{Ford:2004}). Our implementation of OFTI produced this result in 134 minutes, more than an order of magnitude faster than MCMC.

\begin{figure}
\includegraphics[width=0.5\textwidth]{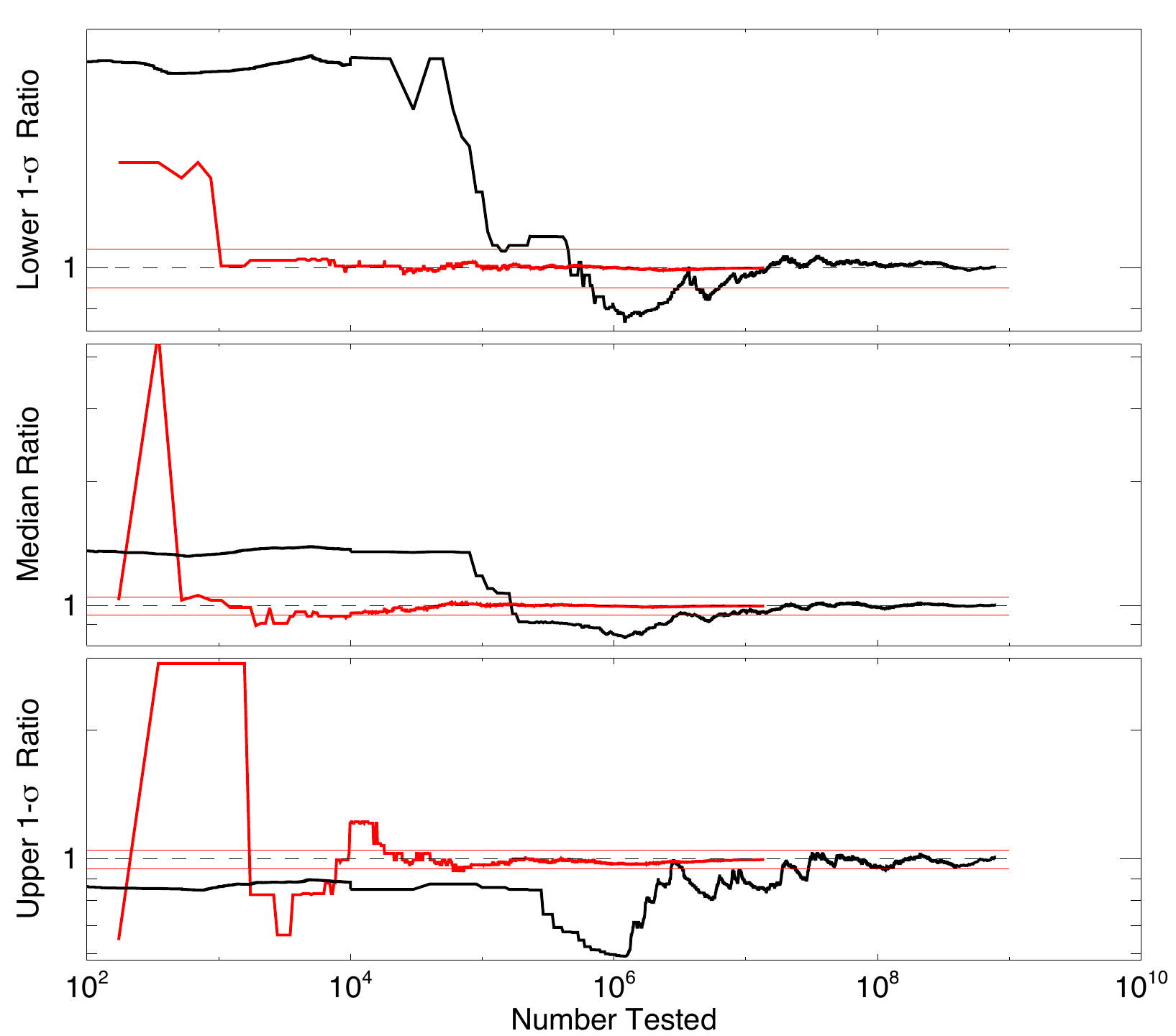}
\caption{Ratios of the partial and final lower 1\,$\sigma$ limit, median, and upper 1\,$\sigma$ limit for one Metropolis-Hastings MCMC chain (black), and one OFTI run (red), computed for all published astrometry of 51 Eri b. OFTI converges on the appropriate median solution after testing approximately $10^4$ sets of orbital elements, while MCMC continues to ``wander'' in a correlated fashion until accepting approximately $10^7$ orbits. Red horizontal lines are located at ratios of 1.05 and 0.95, and a black dashed line is located at a ratio of 1.00.}
\label{fig:changing_medians}
\end{figure}

To demonstrate the differences in the random errors incurred by OFTI and the systematic errors of MCMC, and to illustrate OFTI's computational speed for short orbital arcs, we calculated how the semi-major axis distributions generated by OFTI and MCMC changed as more sets of orbital elements were accepted for OFTI, and generated for MCMC. For both OFTI and MCMC, we calculate the median of the first $n$ semi-major axes tested as a function of $n$, resulting in an array of medians for OFTI and an array of medians for MCMC. We then take the ratio of each number in these arrays to the median of the complete distribution of tested semi-major axes. Since MCMC and OFTI converge on the same distributions, the medians of both complete semi-major axis distributions are the same. As $n$ approaches the total number of orbits tested, the partial distributions approach the complete distribution, and the ratios approach 1. We repeat this procedure for the lower and upper 1\,$\sigma$ limits of the OFTI and MCMC semi-major axis distributions. These results are shown in Figure \ref{fig:changing_medians} for the orbit of 51 Eri b. Note that this represents the number of orbits tested, rather than number of orbits accepted, and so is directly proportional to computation time. After approximately $10^4$ orbits are tested, the OFTI semi-major axis distribution (red line) converges on the final median semi-major axis (to within 5\,\%), while the MCMC semi-major axis distribution (black line) suffers from systematic over- and under-estimates of the final semi-major axis value until more than 3 orders of magnitude more orbits have been tested. Similarly, OFTI converges on the appropriate 1\,$\sigma$ upper and lower limits for the output semi-major axis distribution (to within 5\,\%) after approximately $10^5$ orbits are tested, while it takes MCMC $10^8$ correlated steps in order to do the same. 

\begin{figure}
\includegraphics[width=0.5\textwidth]{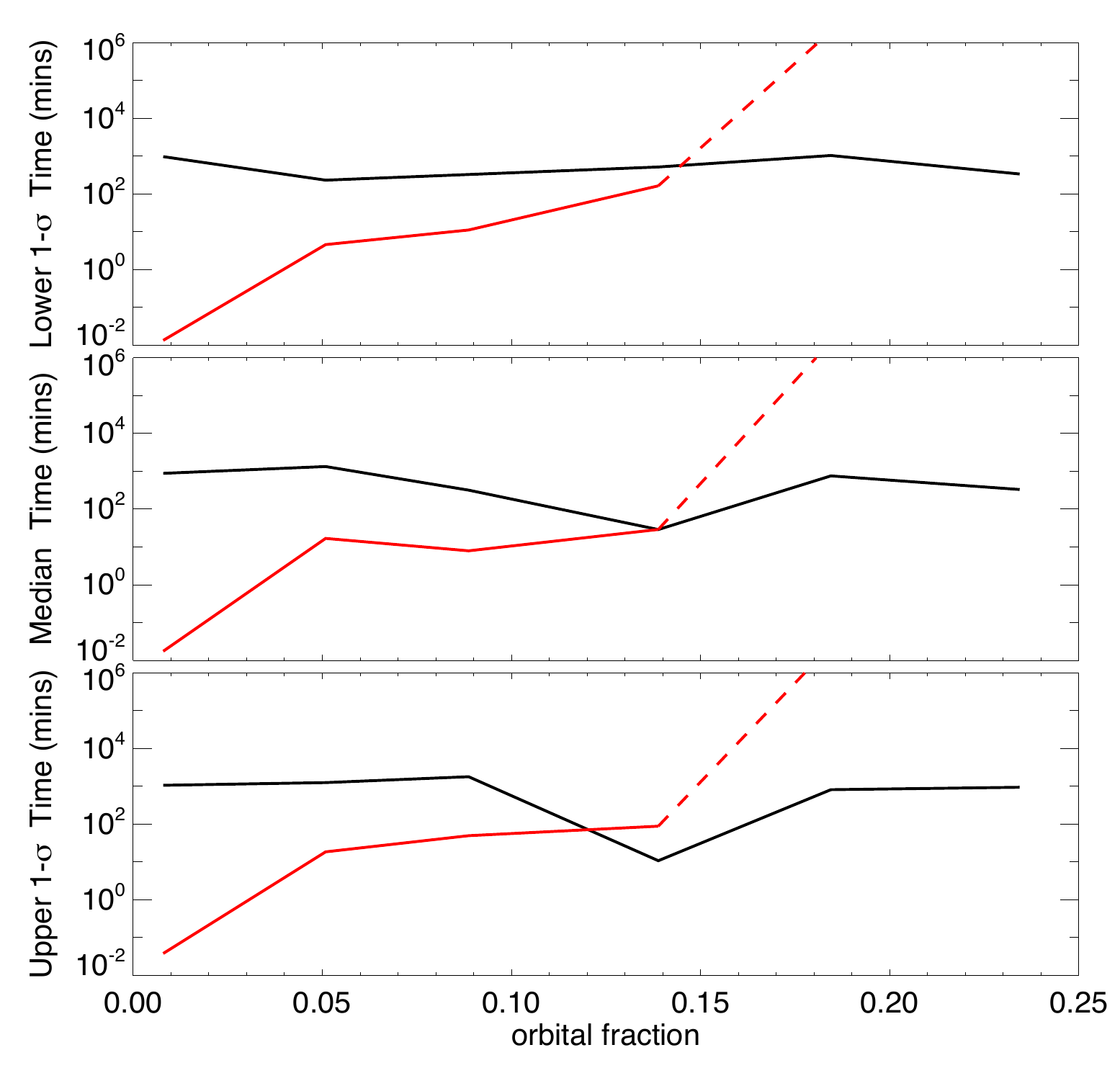}
\caption{Runtime per CPU needed for OFTI (red) and MCMC (black) to determine (within 1$\%\,$ error) the lower 1\,$\sigma$ limit, median, and upper 1\,$\sigma$ limit of the complete $a$ distribution, as a function of the orbital fraction covered by input astrometry of $\beta$ Pic b. As orbital fraction decreases, OFTI performance improves, while MCMC performance slightly improves. For orbital fractions greater than 15\,\%, we extrapolated the depicted behavior from the time OFTI took to accept 50 orbits.}
\label{fig:timeplot}
\end{figure}

OFTI is most efficient for astrometry covering smaller fractions of orbits, while MCMC achieves convergence faster for larger fractions of orbits. Figure \ref{fig:timeplot} illustrates this difference by displaying the wallclock time per CPU needed for each method to achieve convergence using astrometry from $\beta$ Pic b \citep{Millar-Blanchaer:2015}. In order to compare the time for convergence by MCMC and OFTI, we define a proxy for convergence time: our distributions are said to be ``converged'' for a statistic of interest (e.g. $a$ median, $i$ 68\,\% confidence interval) when the statistic is within 1$\,\%$ of the final value, where the final value is calculated from a distribution of $10^4$ accepted orbits.

\begin{figure}
\includegraphics[width=0.5\textwidth]{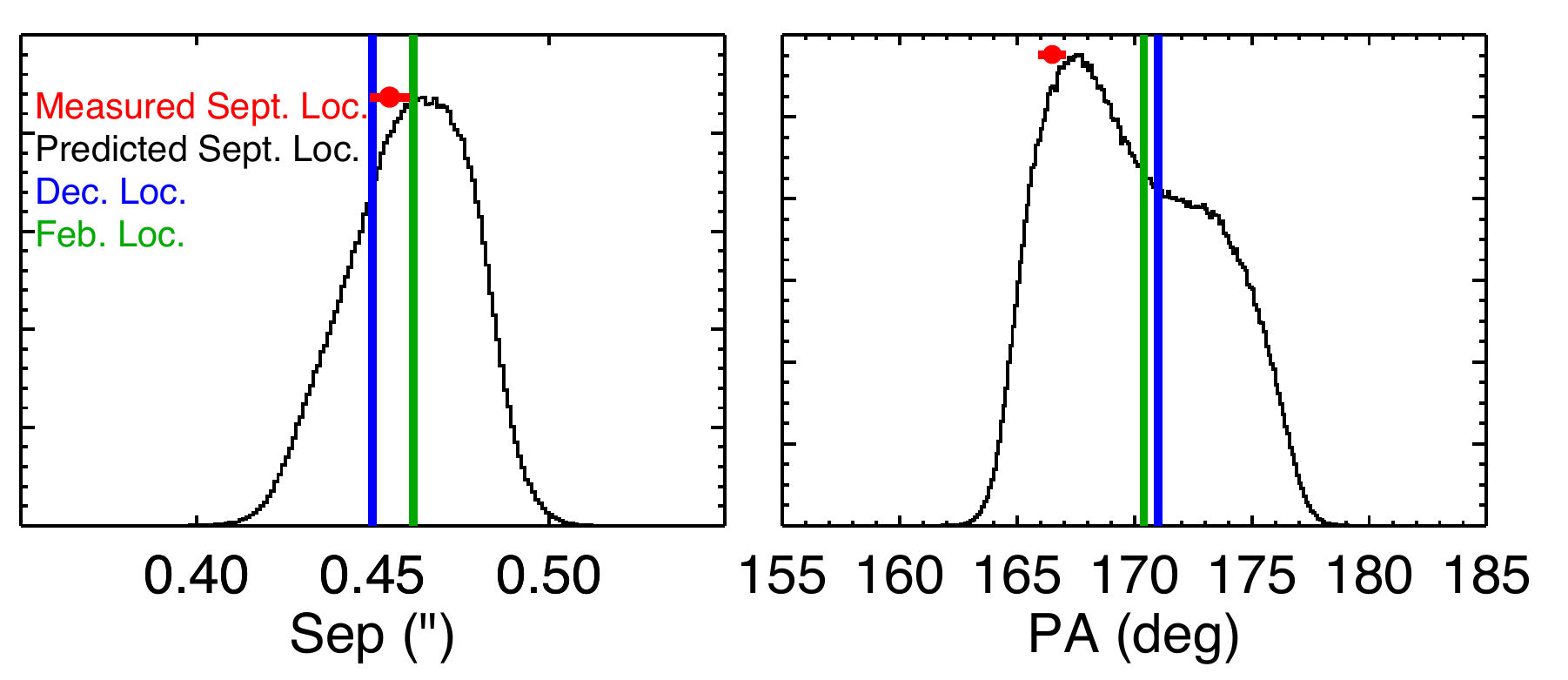}
\caption{PDFs in separation and position angle (black curves) predicting the location of 51 Eri b on 2015 September 15, given all previously known epochs of astrometry. The vertical blue lines show the astrometric location of 51 Eri b on 2015 February 1, the green lines the location on 2014 December 18, and the red horizontal bars the measured $\rho$, $\theta$.}
\label{fig:goi2}
\end{figure}

For astrometry covering small fractions of a total orbit, OFTI can compute accurate future location predictions in much less time than MCMC. We illustrate this application in Figure \ref{fig:goi2}, which shows probability distributions predicting the $\rho$, $\theta$ of 51 Eri b on 2015 September 15 from four earlier astrometric points taken over a timespan of less than 2 months \citep{Macintosh:2015}. Overplotted is the actual measured location of 51 Eri b. The predicted and observed medians for $\rho$ are $0.455\pm\,0.086$ and $0.4547\pm\,0.0057$, and for $\theta$ are $170\pm\,4$ and $166.5\pm\,0.6$. This prediction analysis was made before the most recent astrometric data were obtained. 51 Eri A was unobservable between February and September of 2015. It took OFTI less than 5 minutes (running in parallel on ten 2.3\,GHz AMD Opteron 6378 processors) to produce these predictions. 

\subsection{Estimate of Performances}

OFTI is a rejection-sampling method, meaning that it works by randomly sampling the parameter space of interest, then rejecting the sampled areas that do not match the data. As astrometry drawn from a larger fraction of the orbit becomes available, the orbit becomes more constrained, and the areas of parameter space that match the data shrink, so that OFTI becomes less efficient. A useful analogy for this phenomenon is throwing darts at a dartboard: when astrometry from only a small fraction of an orbit is available, many diverse orbits might fit the data, and a large fraction of the dartboard is acceptable, which results in a high acceptance rate. However, when more astrometry becomes available, a much smaller set of orbits will fit the data, and a much smaller fraction of the dartboard is acceptable, resulting in a lower acceptance rate.

Accordingly, OFTI is most efficient for astrometry covering a short fraction of an orbit, typically less than 15\,\% of the full orbital period. Because directly imaged exoplanets and brown dwarfs have large physical separations from their primary objects (greater than several au), OFTI is ideal for fitting the orbits of directly imaged systems, especially when the time spanned by direct imaging observations is short. 

OFTI is also optimal when a quick estimate of the mean of a distribution is required. This will be particularly helpful in planning follow-up observations for space missions, as it allows to quickly estimate the optimal time for observations, also taking into account the possibility of the planet passing behind the star (see e.g. \citealt{Savransky:2015}). As Figure \ref{fig:changing_medians} shows, OFTI can converge on an estimate for the median of the 51 Eri b semi-major axis distribution within 5\,\% of the true median after fewer than $10^4$ orbits are tested. While an implementation of MCMC would have to run to completion in order to avoid a biased estimate of the semi-major axis distribution, the independence of successive OFTI trials allows OFTI to converge on an unbiased estimate much faster.

\section{Applications}

We use OFTI to fit orbits to 10 sets of astrometry from directly imaged exoplanets, brown dwarfs, and low-mass stars in the literature. Each substellar object has at least two published epochs of astrometry. We chose mostly objects for which an orbital fit has not been calculated because the available astrometry covers a short fraction of the object's orbit. In performing these fits, we make the natural assumption that all objects are bound, and that all objects execute Keplerian orbits, as the chance of catching a common proper motion companion in the process of ejection or during the closest approach of two unassociated objects is particularly small.

We calculate fits using only the data available in the literature. Random and systematic errors in the astrometry available in the literature can bias these results. In particular, systematic errors in the measurement of plate scale or true north of the various instruments used to compile a single astrometric data set can significantly change orbit fits. Sharp apparent motion due to astrometric errors is likely to be fit as a higher eccentricity orbit; more astrometric data are needed to identify outliers of this nature.

For each substellar object, we compiled relative astrometry, distances, and individual object mass estimates from the literature (see Appendix). Values of the companion mass with error bars were given for 2M~1207~b, $\kappa$~And~B, CD-35~2722~B, and GJ~504~b, and these were adopted as reported by the listed references. For HD~1160~B and C, $\eta$~Tel~B, and HR~3549~B, no central value with error bars were given, but instead a range of masses was provided; in these cases we adopted the middle of the range. We note that for visual orbits mass of the system enters Kepler's third law as the sum of the mass of both components, and so uncertainties in the orbit are dominated by uncertainties in the primary mass. We used Gaussian mass priors centered at the sum of the appropriate primary and secondary masses, with FWHM conservatively chosen to be the sum of the two mass uncertainties. For companion masses less than $15\,\%$ of the primary mass, which was the case for all objects except HIP~79797 B and 2M~1207, we neglected the uncertainty in the companion mass, and simply adopted the uncertainty of the primary. We used symmetric Gaussian priors in parallax or distance (we used distance priors only if no parallax was available in the literature), and symmetric or asymmetric Gaussian priors (asymmetric where asymmetric error bars were given) in total mass. Asymmetric Gaussian distributions consist of two half-Gaussians with individual $\sigma$ values, pieced together at the median of the aggregate distribution.

For each orbit, we provide:

\begin{itemize}
\item A table listing the maximum probability (maximum product of $\chi ^2$ probability and prior probabilities), minimum $\chi ^2$, median, 68\,\% confidence interval, and 95\,\% confidence interval orbital elements
\item A triangle plot showing posterior distributions for each orbital element and 2-dimensional covariances for each pair of orbital elements
\item A 3-panel plot showing 100 orbits drawn from the posterior distributions
\end{itemize}

\subsection{GJ~504}

\begin{figure}
\includegraphics[width=0.5\textwidth]{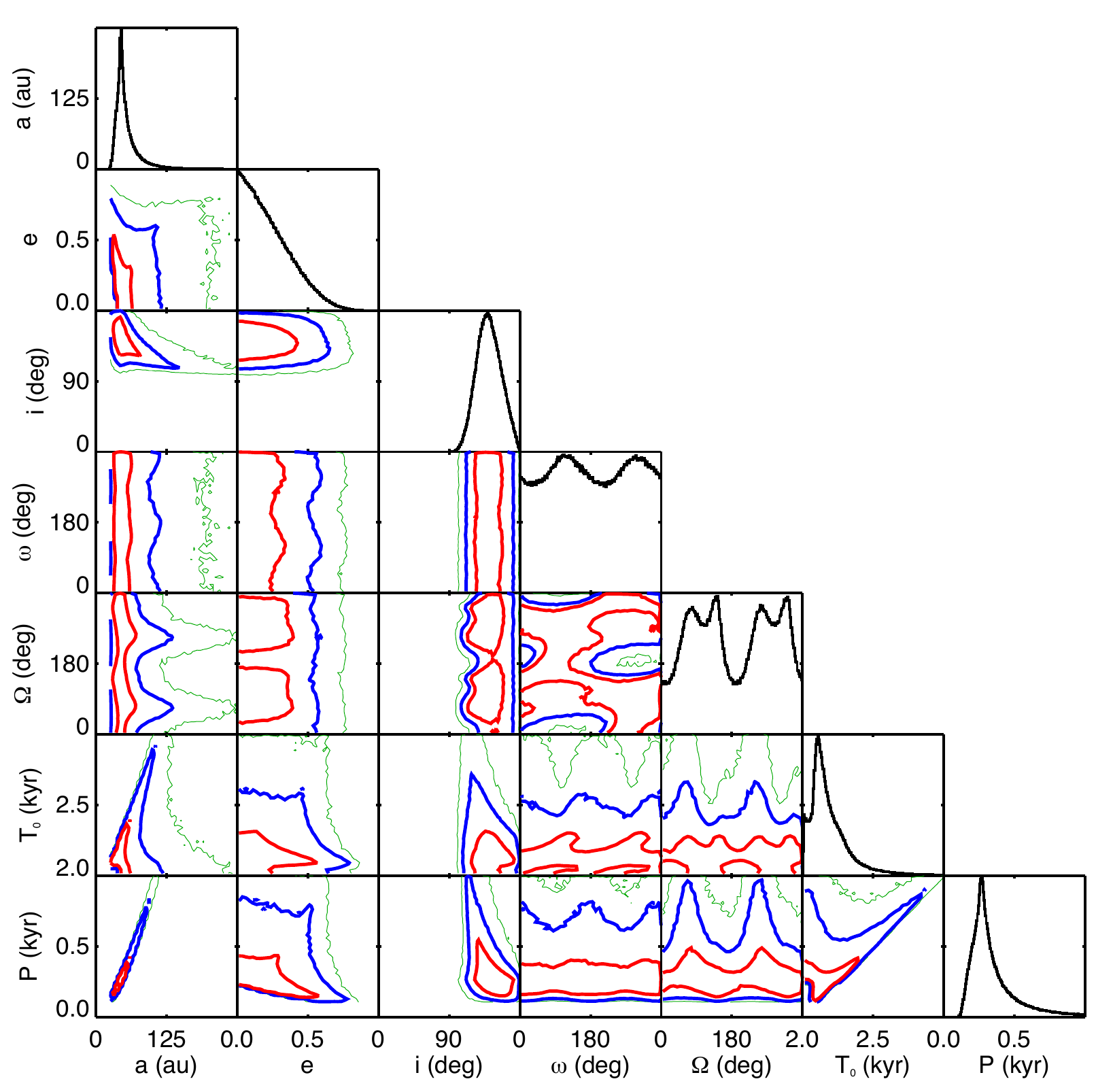}
\caption{Posterior PDFs for the orbit of GJ~504~b with respect to GJ~504~A. Black lined panels on the diagonal show the posterior probability distributions for each of the orbital parameters, while off-diagonal plots show  two-dimensional covariance contour plots. Red lines depict 68\,\% contours, blue lines depict 95\,\% contours, and green lines depict 99\,\% contours.}
\label{fig:gj504_covariance}
\end{figure}

\begin{figure}
    \centering
    \includegraphics[width=0.45\textwidth]{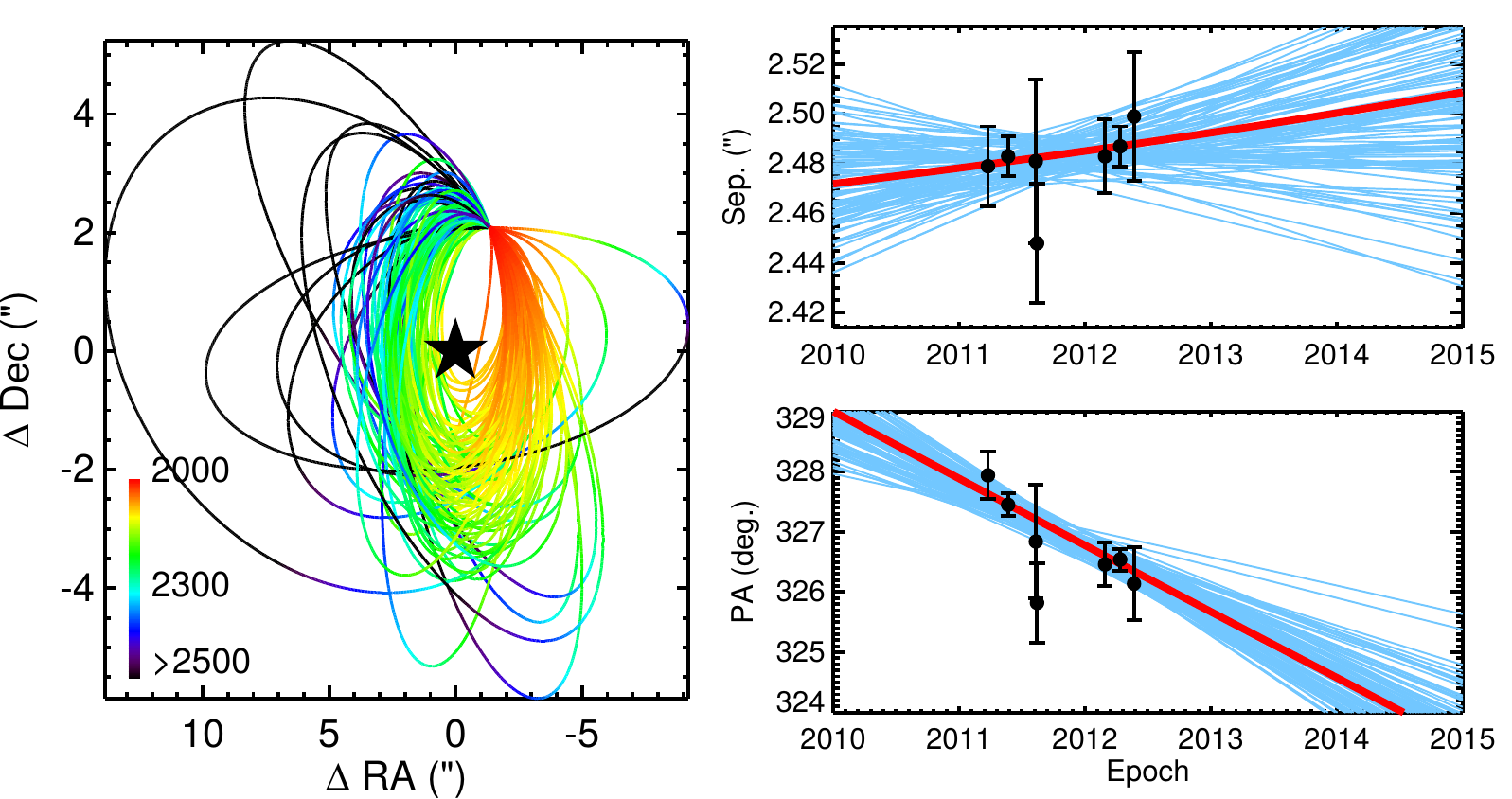}
    \caption{100 orbits fit to relative astrometry of GJ~504~b with respect to GJ~504~A, randomly selected from OFTI posterior PDFs. Left: orbital motion of GJ~504~b with respect to GJ~504~A over an orbital period. As elapsed time since the most recent observational epoch increases, the color of the orbital track changes from red to green to black (see colorbar in lower left corner). The black star indicates the primary. Right: relative separations (top) and position angles (bottom), together with the observed measurements and errors (black points with error bars), and minimum $\chi ^2$ orbit (red line). These orbits are the same as the orbits plotted in the left panel.}
    \label{fig:gj504_orbit}
\end{figure}

GJ~504~b is the coldest and bluest directly-imaged exoplanet to date, and one of the lowest mass. Its discovery was reported by \citet{Kuzuhara:2013}, who also perform a rejection-sampling orbit fit similar to OFTI \citep{Janson:2011}. Masses and astrometry are provided in Tables \ref{tab:starparams} and \ref{tab:GJ504}, and results are shown in Table \ref{tab:gj504_outputs} and Figures \ref{fig:gj504_covariance} and \ref{fig:gj504_orbit}. Our results are consistent with the posterior distributions they find (noting that \citet{Kuzuhara:2013} have used a flat prior in eccentricity). Using a linearly descending prior in eccentricity, we find a median semi-major axis of 48\,au, with 68\,\% confidence between 39 and 69\,au, and a corresponding period of 299 years, with 68\,\% confidence between 218 and 523 years. We also note that \citet{Kuzuhara:2013} find an $e$ posterior that decreases with eccentricity, as we do for OFTI calculations performed assuming both a uniform eccentricity prior and a linearly descending eccentricity prior.

We calculated fraction of orbital coverage by dividing the time spanned by observations by the 68\,\% confidence limits of the posterior distribution in period produced by OFTI. The calculated orbital fraction for the orbit of GJ~504~b is $0.4\,^{+0.1}_{-0.2}$\,\%.

\begin{figure}
\includegraphics[width=0.5\textwidth]{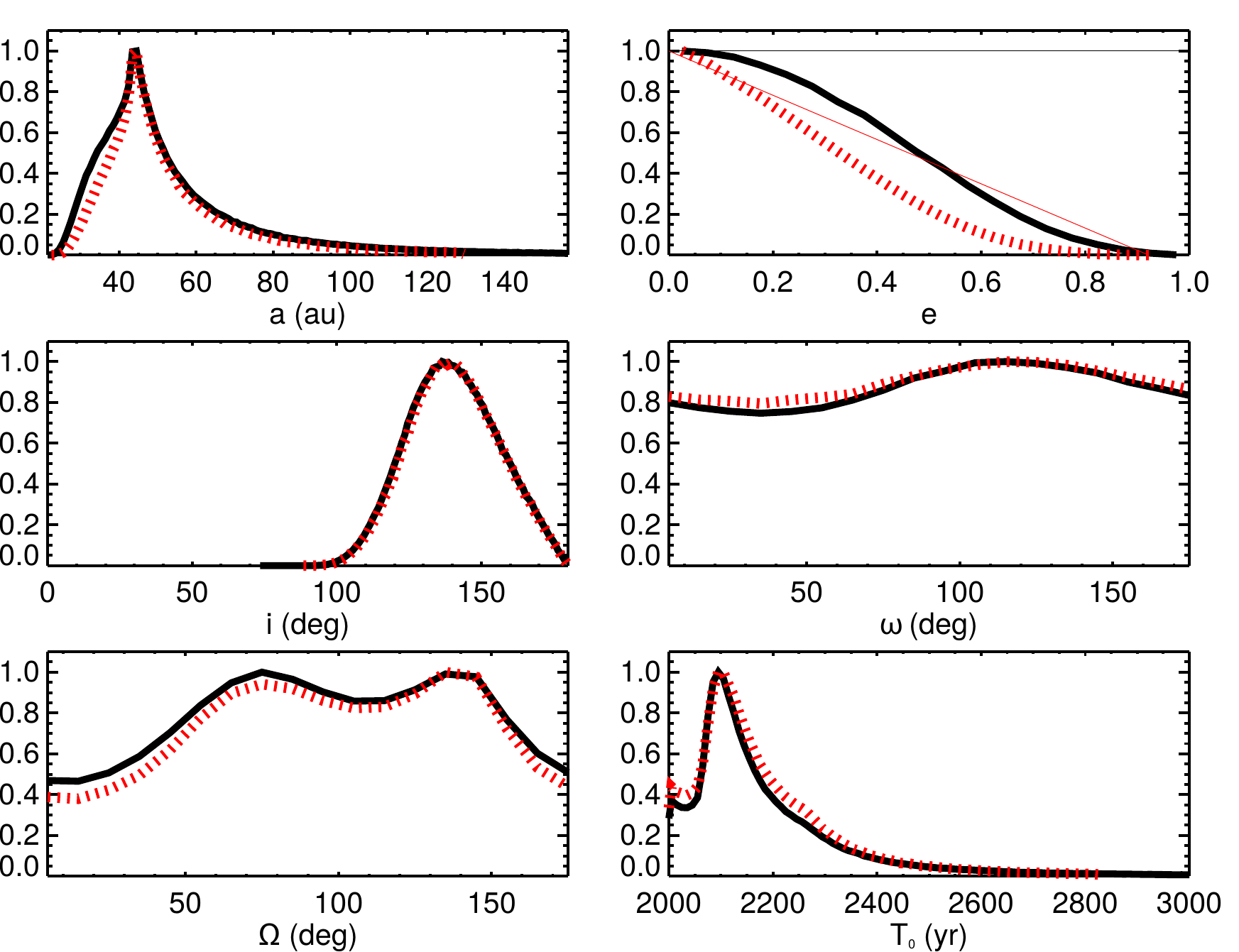}
\caption{Posterior PDFs for the orbit of GJ~504~b, calculated using a linearly descending prior in eccentricity (black), and a uniform prior in eccentricity (red). The choice of prior strongly affects the eccentricity posterior, but the other orbital element posteriors are generally the same. The largest change seen is in semi-major axis, where the distribution shifts from $48^{+22}_{-9}$ au to $47^{+26}_{-11}$ au when changing the linearly descending eccentricity prior to a uniform one. The two relevant eccentricity priors are plotted in the eccentricity panel as thin red and black lines, but the black prior is not visible behind the black posterior.}
\label{fig:prior_effect}
\end{figure}

To illustrate the impact of our choice of eccentricity prior on the results, we performed another fit to the astrometry of GJ~504~b using a uniform, rather than a linearly descending, prior in eccentricity. The results are shown in Figure \ref{fig:prior_effect}. The use of a different eccentricity prior changes the eccentricity posterior PDF, but does not significantly affect the other posterior PDFs. For example, when a linearly descending eccentricity prior is used, the semi-major axis posterior is $48\,^{+22}_{-9}$\,au, and when a uniform eccentricity prior is used, the posterior shifts to $47\,^{+26}_{-11}$\,au.

\begin{deluxetable*}{ccccccc}
\tabletypesize{\scriptsize}
\tablecaption{Orbit of GJ~504~b with respect to GJ~504~A}
\tablewidth{0pt}
\tablehead{
\colhead{orbital element} & \colhead{unit} & \colhead{max probability} & \colhead{min $\chi ^2$} & \colhead{median} & \colhead{68\% confidence range} & \colhead{95\% confidence range}
}
\startdata
$a$ & au &44.48& 67.24& 48& 39-69& 31-129\\
$P$ & yr &268.56& 508.23& 299& 218-523& 155-1332\\
$e$ & & 0.0151& 0.1519& 0.19& 0.05-0.40& 0.01-0.62\\
$i$ & $^{\circ}$ &  142.2& 131.7& 140& 125-157& 111-171\\
$\omega$ & $^{\circ}$ & 91.7& 4.9& 95& 31-151& 4-176\\
$\Omega$ & $^{\circ}$ & 133.7& 61.6& 97& 46-146& 8-173\\
$T_0$ & yr & 2228.11& 2419.96& 2145.10& 2068.06-2310.13& 2005.07-2825.03\\
\label{tab:gj504_outputs}
\tablenotetext{}{Note: $\Omega$ and $\omega$ have been wrapped between 0 and 180$^{\circ}$, and $T_0$ has been wrapped between 1995 and 1995 + 1 period. The maximum probability orbit was calculated by multiplying the output $\chi ^2$ likelihood by the priors, and taking the orbit with the maximum product. The acceptance rate was 0.13\,\%.}
\end{deluxetable*}

\subsection{HD~1160}

HD~1160~A hosts two known companions: HD~1160~B, a low-mass object at the stellar/brown dwarf boundary \citep{Garcia:2016}, and HD~1160~C, an M3.5 star at a wider projected separation than HD~1160~B \citep{Nielsen:2012}. We compute fits to both the orbits of HD~1160~B and HD~1160~C with respect to HD~1160~A based on astrometry compiled in Tables \ref{tab:hd1160B} and \ref{tab:hd1160C}, and mass and distance values provided in Table \ref{tab:starparams} in the Appendix. Preliminary orbital fits were provided by \citet{Nielsen:2013}, but an updated fit including the latest astrometry published in \citet{Maire:2016} has not been computed. The results of our fits are shown in Tables \ref{tab:hd1160b_outputs} and \ref{tab:hd1160c_outputs} and Figures \ref{fig:hd1160b_covariance} - \ref{fig:hd1160c_orbit}. Because HD~1160~B and C are non-planetary companions, we used a uniform prior in $e$, rather than the empirically derived linearly descending prior used for exoplanets. This choice is supported by evidence that the empirical eccentricity distribution of long-period stellar binaries is approximately uniform (e.g. \citealt{Duchene:2013}). 

For HD~1160~B, the fraction of orbital coverage is $2.5\,^{+2.2}_{-1.8}$\,\%, while for HD~1160~C, the fraction of orbital coverage is $0.11\,\pm0.06$\,\%. 

OFTI constrains the range of possible inclination angles for the orbit of HD~1160~B tightly, with a 68\,\% confidence interval of 96-119\,$^{\circ}$, indicating close to an edge-on orbit. We find a semi-major axis of $77\,^{+96}_{-27}$\,au, and a period of $479\,_{-227}^{+1148}$ years. A high eccentricity is favored. As seen in Figure \ref{fig:hd1160b_orbit}, the minimum $\chi ^2$ orbit passes through all 1\,$\sigma$ error bars in $\rho$ and $\theta$, allowing OFTI to converge on a set of permissible orbits in fewer iterations than, for example, the orbit of $\eta$~Tel~B, whose data set contains several clear outliers (see Section 3.6). 

The fit favors a more face-on orbit for HD~1160~C than for HD~1160~B, returning an inclination angle of $146\,^{+17\, \circ}_{-18}$. The probability that the inclination angle of HD~1160~B is within 10\,$^{\circ}$ of the inclination angle of HD~1160~C is 8\,\%. It is fairly typical for triple stellar systems to be non-coplanar (e.g. \citealt{Fekel:1981}, \citealt{Tokovinin:2004}), so this result is plausible. In keeping with the larger projected separation of HD~1160~C, we find a larger semi-major axis, $651\,_{-160}^{+432}$\,au, and period, $11,260\,^{+12,930}_{-3,898}$ years, for HD~1160~C compared to HD~1160~B. 

\subsection{HIP~79797}

HIP~79797~Ba and HIP~79797~Bb are a close binary brown dwarf system orbiting the A star HIP~79797~A, first detected as an unresolved companion by \citet{Huelamo:2010}, and resolved into a binary by \citet{Nielsen:2013}. \citet{Nielsen:2013} perform a preliminary orbit fit using MCMC, but note that the MCMC fit was un-converged. Masses and astrometry are provided in Tables \ref{tab:starparams} and \ref{tab:hip79797}. Results are shown in Table \ref{tab:79797_outputs} and Figures \ref{fig:hip79797_covplot} and \ref{fig:hip79797_orbit}. Like \citet{Nielsen:2013}, our OFTI fit favors an edge-on orbit ($i=83\,^{+10 \, \circ}_{-17}$), but we find a significantly smaller semi-major axis ($3\,^{+5}_{-1}$\,au, where Nielsen et al give a semi-major axis of $25\,^{+97}_{-19}$\,au) and corresponding period ($23\,^{+63}_{-12}$ years, where \citet{Nielsen:2013} give a period of $380\,^{+3700}_{-330}$ years). Our fit favors high eccentricity orbits. Again, we use a uniform prior in eccentricity. The calculated orbital fraction for the orbit of HIP~79797~Ba and HIP~79797~Bb is $10\,^{+11}_{-8}$\,\%. 

\subsection{HR~3549}

HR~3549~B is a brown dwarf orbiting the A0V star HR~3549~A, discovered by \citet{Mawet:2015}, and followed up by \citet{Mesa:2016}. Masses and astrometry are provided in Tables \ref{tab:starparams} and \ref{tab:hr3549}. \citet{Mesa:2016} use a LSMC technique to constrain the orbit of HR~3549, and provide orbital elements for three orbits with greatest likelihood (least $\chi ^2$). While minimization algorithms like these are effective at finding local maxima in likelihood space, OFTI produces full Bayesian posteriors that show probability, rather than likelihood. Our OFTI fit, in contrast to that of \citet{Mesa:2016}, makes use of both astrometry points \citet{Mesa:2016} provide, rather than just one, and uses a Gaussian prior in mass with full width at half maximum (FWHM) equal to 5\,\% of the reported value in \citet{Mesa:2016}, rather than a fixed value. Our results are provided in Table \ref{tab:hr3549_outputs} and Figures \ref{fig:hr3549_covplot} and \ref{fig:hr3549_orbit}. We find a semi-major axis of $94\,^{+65}_{-28}$\,au, which contains one of the semi-major axis values reported by \citet{Mesa:2016} (133.2\,au), while the other two semi-major axis values \citet{Mesa:2016} report (299.7 and 441.2\,au) are at 1.7 and 2.9\,$\sigma$, respectively, of our semi-major axis distribution. Like \citet{Mesa:2016}, we find that most values of eccentricity, and values of $i>90\,^\circ$, are consistent with the astrometry, a reflection of the fact that the astrometry covers only a fraction of the full orbital arc ($0.5\,^{+0.4}_{-0.3}$\,\%).

\subsection{2M~1207}

2MASSW J1207334-393254 b (hereafter 2M~1207~b) is a planetary mass companion orbiting 2M~1207~A, a brown dwarf with mass approximately three times that of 2M~1207~b. 2M~1207~b was first discovered using the Very Large Telescope with NACO by \citet{Chauvin:2004}. It was followed up first by \citet{Song:2006}, who confirmed the common proper motion of the pair, establishing 2M~1207~b as a bound companion, and again by \citet{Mohanty:2007}. Masses and astrometry are provided in Tables \ref{tab:starparams} and \ref{tab:2m1207}, and orbital fits are shown in Table \ref{tab:2m1207_outputs} and Figures \ref{fig:2m1207_covariance} and \ref{fig:2m1207_orbit}. Using a linearly descending prior in eccentricity, we find a median semi-major axis of 46\,au, with 68\,\% confidence between 31 and 84\,au. We only loosely constrain the range of possible inclination angles, and determine that high eccentricity ($>0.8$) orbits are dispreferred. The median orbital period is 1,782 years, with 95\,\% confidence between 633 years and 20,046 years. The calculated orbital fraction for the orbit of 2M~1207~b is $0.06\,^{+0.05}_{-0.03}$\,\%.

\subsection{$\kappa$~And}

$\kappa$~And~B is a substellar companion orbiting the late B star $\kappa$~And. The discovery of $\kappa$~And~B was first reported by \citet{Carson:2013}, and followed up by \citet{Bonnefoy:2014}. Masses and astrometry are provided in Tables \ref{tab:starparams} and \ref{tab:kand}, and results are shown in Table \ref{tab:kapAnd_outputs} and Figures \ref{fig:kapand_covariance} and \ref{fig:kapand_orbit}. Using a linearly descending prior in eccentricity, we find a median semi-major axis of 77\,au, with 68\,\% confidence between 54 and 123\,au. Eccentricity remains mostly unconstrained after our analysis, but inclination is determined to be between 59\,$^{\circ}$ and 159\,$^{\circ}$ with 95\,\% confidence. The median orbital period is 378 years, with 95\,\% confidence between 144 years and 2,033 years. The calculated orbital fraction for the orbit of $\kappa$~And~B is $0.2\,^{+0.2}_{-0.1}$\,\%.

\subsection{$\eta$~Tel}

$\eta$~Tel~B is a brown dwarf orbiting the A0V star $\eta$~Tel~A (\citealt{Lowrance:2000}, \citealt{Neuhauser:2011}). Masses and astrometry are provided in Tables \ref{tab:starparams} and \ref{tab:etaTel}, and results are shown in Table \ref{tab:etaTel_outputs} and Figures \ref{fig:etatel_covariance} and \ref{fig:etaTel_orbit}. We find a median semi-major axis of 192 au, with 68\,\% confidence between 125 and 432\,au. The corresponding median period is 1,493 years, with 68\,\% confidence between 781 and 5,028 years. OFTI produced a well-constrained posterior in inclination angle, with 95\,\% of orbits having an inclination angle between 40$\,^{\circ}$ and 120$\,^{\circ}$. Several data points are outliers, which results in a comparatively low orbital acceptance rate of 0.002\,\%. We use a uniform prior in eccentricity. The calculated orbital fraction for the orbit of $\eta$~Tel~B is $0.7\,^{+0.7}_{-0.5}$\,\%.

\subsection{2M~0103-55}

2MASS~J01033563-5515561~(AB)~b (hereafter 2M~0103-55~(AB)~b) is a 12-14 Jupiter-mass object in orbit with respect to 2M~0103-55~(AB), a binary consisting of two low mass stars \citep{Delorme:2013}. \citet{Delorme:2013} first reported the discovery of 2M~0103-55~(AB)~b, and provide two epochs of astrometry. Masses and astrometry are provided in Tables \ref{tab:starparams} and \ref{tab:2M0103}, and results are shown in Table \ref{tab:2m0103_outputs} and Figures \ref{fig:2m0103_covariance} and \ref{fig:2m0103_orbit}. Two epochs of astrometry is generally too short of a baseline for the two implementations of MCMC discussed in this work to converge within the timescale of a few days, but OFTI quickly returns the appropriate posteriors. Using a linearly descending prior in eccentricity, we find a semi-major axis of 102\,au, with 68\,\% confidence between 75 and 149\,au, and a corresponding period of 1,682 years, with 68\,\% confidence between 1,054 and 2,990 years. OFTI also returns a firm lower limit on inclination angle, with 95\,\% of orbits having inclination angles greater than 112$\,^{\circ}$. The calculated orbital fraction for the orbit of 2M~0103~(AB)~b is $0.6\,^{+0.4}_{-0.3}$\,\%.

\subsection{CD-35~2722}

CD-35~2722~B is an L-dwarf companion to the M1 dwarf CD-35~2722~A, discovered by the Gemini NICI planet-finding campaign \citep{Wahhaj:2011}. \citep{Wahhaj:2011} report two epochs of relative astrometry and show that CD-35~2722~B is a bound companion on the basis of common proper motion. Masses and astrometry are provided in Tables \ref{tab:starparams} and \ref{tab:CD35}, and results are shown in Table \ref{tab:cd35_outputs} and Figures \ref{fig:cd35_covariance} and \ref{fig:2m0103_orbit}. With OFTI, we find a median semi-major axis of 115\,au, with 68\,\% confidence between 74 and 216\,au. The corresponding period is 1,853 years, with 68\,\% confidence between 947 and 4,772 years. Inclination angle and eccentricity remain mostly unconstrained. We use a uniform prior in eccentricity. The calculated orbital fraction for the orbit of CD-35~2722~B is $0.05\,^{+0.05}_{-0.03}$\,\%.

\section{Conclusions}

OFTI is a novel orbit-fitting method that reproduces the outputs of MCMC in orders of magnitude less time when fitting astrometric data covering only small fractions of orbits. A key difference with MCMC is that each OFTI orbit is independent of the others, whereas an MCMC chain produces a series of correlated values which only define the posterior PDF once the chains have fully converged. This difference makes OFTI an ideal tool when parameter estimates are required quickly, as in the context of a space mission. For example, when planning future observations, OFTI can quickly compute the expected decrease in errors on orbital parameters for different observing cadences without having to wait for multiple MCMC chains to converge.

In this paper, we have demonstrated the accuracy of the OFTI method by comparing the outputs of our implementation of OFTI with those of MCMC, the speed of OFTI by analyzing the outputs produced for varying input orbital fraction, and the utility of OFTI by providing fits to ten sets of astrometric data covering very short orbital arcs ($<3\,\%$ orbital coverage for all but one) from the literature.

OFTI is a useful tool for constraining the orbital parameters of directly imaged long-period exoplanets, brown dwarfs, and long-period stellar binaries. It has been applied to fitting the orbits of exoplanets imaged by the GPIES campaign and extremely long-period brown dwarfs, and will be used to fit the orbits of future \textit{WFIRST} targets. OFTI's efficiency will be critical for a space-based mission like \textit{WFIRST}, allowing future observations to be planned effectively and efficiently.

\acknowledgments

We thank the anonymous referee for helpful comments that improved the quality of this work. S.B. has been supported by the National Science Foundation Research Experiences for Undergraduates Program under Grant No. AST-1359346 and the Stanford Summer Research Early-Identification Program. S.B. would also like to acknowledge and thank Charles Royce and the Royce Fellowship for their support, and Jasmine Garani for useful discussion. Thanks to Mike Fitzgerald, Anand Sivaramakrishnan, Alexandra Greenbaum, Max Millar-Blanchaer, and Vanessa Bailey for helpful comments. Based on observations obtained at the Gemini Observatory, which is operated by the Association of Universities for Research in Astronomy, Inc., under a cooperative agreement with the National Science Foundation (NSF) on behalf of the Gemini partnership: the NSF (United States), the National Research Council (Canada), CONICYT (Chile), the Australian Research Council (Australia), Minist{\'e}rio da Ci{\^e}ncia, Tecnologiae Inovacao (Brazil) and Ministerio de Ciencia, Tecnolog{\'i}a e Innovaci{\'o}n Productiva (Argentina). E.L.N., S.B., B.M., F.M., and M.P. were supported by NASA grant NNX14AJ80G. R.J.D.R, D.R, J.J.W, J.R.G have been supported by NSF grant AST-1518332, National Aeronautics and Space Administration (NASA) Origins grant NNX15AC89G, and NASA NExSS grant NNX15AD95G. This work benefited from NASA’s Nexus for Exoplanet System Science (NExSS) research coordination network sponsored by NASA’s Science Mission Directorate.

{}

\appendix

\newpage

\clearpage
\begin{landscape}
\begin{deluxetable*}{ccccccc}
\tabletypesize{\scriptsize}
\tablecaption{Relevant Distance \& Mass Parameters}
\tablewidth{0pt}
\tablehead{
\colhead{Orbit [primary, companion]} & \colhead{Parallax (mas)} & \colhead{Primary Mass ($M_{\odot}$)} & \colhead{Secondary Mass ($M_J$)} & \colhead{Mass Refs.} & \colhead{OFTI Input Mass ($M_{\odot}$)} & \colhead{Orbit Fraction$^{d}$}
}
\startdata
HD~1160~A, HD~1160~B & $9.66\,\pm0.45$ & $1.95\,\pm0.10^{a}$ &  $29$ & \citet{Siess:2000} & $1.95\,\pm0.10$ & $2.5\,^{+2.2}_{-1.8}$\,\% \\
& & & & \citet{Garcia:2016} \\
\\
HD~1160~A, HD~1160~C & $9.66\,\pm0.45$ & $1.95\,\pm0.10^{a}$ & $187$ & \citet{Garcia:2016}, & $2.13\,\pm0.10$ & $0.11\,\pm0.06$\,\%\\
& & & & \citet{Nielsen:2012}  \\
\\
HIP~79797~Ba, HIP~79797~Bb & $19.15\,\pm0.42$ & $0.06\,\pm0.02$, $0.05\,\pm0.02$ & & \citet{Nielsen:2013} & $0.11\,\pm0.04$ & $10\,^{+11}_{-8}$\,\%\\
\\
HR~3549~A, HR~3549~B & $10.82\,\pm0.27$ & $2.30\,\pm0.12^{c}$ & $45$ & \citet{Mesa:2016} & $2.34\,\pm0.12$ & $0.5\,^{+0.4}_{-0.3}$\,\%\\
\\
2M~1207~A, 2M~1207~b& $18.51\,\pm1.03$ & $0.024\,\pm0.005$ & $8\,\pm2$ & \citet{Patience:2012} & $0.032\,\pm0.007$ & $0.06\,^{+0.05}_{-0.03}$\,\%\\
\\
$\kappa$~And~A, $\kappa$~And~B & $19.37\,\pm0.19$ & $2.8\,^{+0.1}_{-0.2}$ &  $50\,^{+16}_{-13}$ & \citet{Hinkley:2013} & $2.8\,^{+0.1}_{-0.2}$ & $0.2\,^{+0.2}_{-0.1}$\,\%\\
\\
$\eta$~Tel~A, $\eta$~Tel~B & $20.74\,\pm0.21$ & $2.90\,\pm0.15^{b}$ & $35$ & \citet{Wyatt:2007}, & $2.93\,\pm0.15$ & $0.7\,^{+0.7}_{-0.5}$\,\%\\
& & & & \citet{Neuhauser:2011} & \\
\\
2M~0103-55~(AB), 2M~0103-55~(AB)~b & $21.18\,\pm1.37$ & $0.36\,\pm0.04$ & $13$ & \citet{Delorme:2013} & $0.37\,\pm0.04$ & $0.6\,^{+0.4}_{-0.3}$\,\%\\
\\
CD-35~2722~A, CD-35~2722~B & $22.7\,\pm1.0$\,pc$^{c}$& $0.40\,\pm0.05$ & $31\,\pm8$ & \citet{Wahhaj:2011} & $0.43\,\pm0.05$ & $0.05\,^{+0.05}_{-0.03}$\,\%\\
\\
GJ~504~A, GJ~504~b & $56.95\,\pm0.26$ & $1.22\,\pm0.08$ & $4\,^{+4.5}_{-1.0}$ & \citet{Kuzuhara:2013} & $1.22\,\pm0.08$ & $0.4\,^{+0.1}_{-0.2}$\,\%\\
\enddata
\tablenotetext{a}{Interpolated from Siess models \citep{Siess:2000} using an age of 0.05\,Gyr, Z=.02, mv = 7.12; 5\,\% error bars assumed.}
\tablenotetext{b}{5\,\% error bars assumed.}
\tablenotetext{c}{\citet{Shkolnik:2012}; Note that this is a trigonometric distance, not a parallax}
\tablenotetext{d}{Fraction of orbit covered by astrometry }
\tablenotetext{}{Note: All parallaxes taken from the Simbad database.}
\tablenotetext{}{Note: Secondary masses without error bars were presented in the papers referenced as mass ranges, rather than PDFs or median values with error bars. We took the median values of these ranges as inputs for OFTI.}
\label{tab:starparams}
\end{deluxetable*}
\clearpage
\end{landscape}

\begin{deluxetable*}{cccc}
\tabletypesize{\scriptsize}
\tablecaption{Previous Observations of HD~1160~B}
\tablewidth{0pt}
\tablehead{
\colhead{Epoch} & \colhead{$\rho$ (")} & \colhead{$\theta$ ($^{\circ}$)} & \colhead{Reference}
}
\startdata
2002.57 & $0.767\,\pm 0.030$ & $246.2\,\pm 1.0$ & \citet{Nielsen:2012}\\
2003.84 & $0.766\,\pm 0.030$ & $245.6\,\pm 1.0$ & \citet{Nielsen:2012}\\
2005.98 & $0.760\,\pm 0.030$ & $244.7\,\pm 1.0$ & \citet{Nielsen:2012}\\
2008.50 & $0.804\,\pm 0.060$ & $245.3\,\pm 2.0$ & \citet{Nielsen:2012}\\
2010.71 & $0.769\,\pm 0.060$ & $242.8\,\pm 2.0$ & \citet{Nielsen:2012}\\
2010.83 & $0.780\,\pm 0.025$ & $244.3\,\pm 0.2$ & \citet{Nielsen:2012}\\
2010.89 & $0.780\,\pm 0.025$ & $244.5\,\pm 0.2$ & \citet{Nielsen:2012}\\
2010.90 & $0.773\,\pm 0.020$ & $244.9\,\pm 0.5$ & \citet{Nielsen:2012}\\
2011.52 & $0.784\,\pm 0.030$ & $244.0\,\pm 1.0$ & \citet{Nielsen:2012}\\
2011.67 & $0.777\,\pm 0.030$ & $244.9\,\pm 1.0$ & \citet{Nielsen:2012}\\
2011.80 & $0.770\,\pm 0.030$ & $244.5\,\pm 0.2$ & \citet{Nielsen:2012}\\
2011.85 & $0.781\,\pm 0.030$ & $244.4\,\pm 1.0$ & \citet{Nielsen:2012}\\
2014.62 & $0.78087\,\pm0.00106$ & $244.25\,\pm0.13$ & \citet{Maire:2016}\\
2014.62 & $0.78097\,\pm0.00047$ & $243.89\,\pm0.21$ & \citet{Maire:2016}\\
\hline
\\
& & Total Change in $\theta$: & 2.31\,$^{\circ}$
\label{tab:hd1160B}
\tablenotetext{}{Note: change in $\theta$ is given as a coarse proxy for amount of orbital motion over the observed epochs.}
\end{deluxetable*}

\begin{deluxetable*}{cccc}
\tabletypesize{\scriptsize}
\tablecaption{Previous Observations of HD~1160~C}
\tablewidth{0pt}
\tablehead{
\colhead{Epoch} & \colhead{$\rho$ (")} & \colhead{$\theta$ ($^{\circ}$)} & \colhead{Reference}
}
\startdata
2002.57 & $5.167\,\pm0.030$ & $351.0\,\pm0.5$ & \citet{Nielsen:2012}\\
2002.74 & $5.167\,\pm0.035$ & $349.8\,\pm0.5$ & \citet{Nielsen:2012}\\
2002.87 & $5.220\,\pm0.035$ & $349.6\,\pm0.5$ & \citet{Nielsen:2012}\\
2003.53 & $5.185\,\pm0.035$ & $349.5\,\pm0.5$ & \citet{Nielsen:2012}\\
2003.64 & $5.180\,\pm0.035$ & $349.7\,\pm0.5$ & \citet{Nielsen:2012}\\
2003.84 & $5.155\,\pm0.030$ & $350.2\,\pm0.5$ & \citet{Nielsen:2012}\\
2003.86 & $5.179\,\pm0.035$ & $349.6\,\pm0.5$ & \citet{Nielsen:2012}\\
2005.98 & $5.145\,\pm0.030$ & $350.4\,\pm0.5$ & \citet{Nielsen:2012}\\
2007.90 & $5.080\,\pm0.070$ & $349.4\,\pm0.5$ & \citet{Nielsen:2012}\\
2008.50 & $5.155\,\pm0.030$ & $349.7\,\pm0.5$ & \citet{Nielsen:2012}\\
2010.71 & $5.146\,\pm0.030$ & $349.4\,\pm0.5$ & \citet{Nielsen:2012}\\
2010.83 & $5.150\,\pm0.025$ & $349.8\,\pm0.2$ & \citet{Nielsen:2012}\\
2010.89 & $5.160\,\pm0.030$ & $349.6\,\pm0.2$ & \citet{Nielsen:2012}\\
2010.90 & $5.137\,\pm0.020$ & $349.9\,\pm0.5$ & \citet{Nielsen:2012}\\
2011.52 & $5.142\,\pm0.030$ & $349.4\,\pm0.5$ & \citet{Nielsen:2012}\\
2011.67 & $5.140\,\pm0.030$ & $349.5\,\pm0.5$ & \citet{Nielsen:2012}\\
2011.76 & $5.119\,\pm0.070$ & $349.2\,\pm0.5$ & \citet{Nielsen:2012}\\
2011.80 & $5.160\,\pm0.030$ & $349.6\,\pm0.2$ & \citet{Nielsen:2012}\\
2011.85 & $5.140\,\pm0.030$ & $349.4\,\pm0.5$ & \citet{Nielsen:2012}\\
2014.62 & $5.14975\,\pm0.00269$ & $349.17\,\pm0.10$ & \citet{Maire:2016}\\
\hline
\\
& & Total Change in $\theta$: & 1.83\,$^{\circ}$
\label{tab:hd1160C}
\end{deluxetable*}

\begin{deluxetable*}{cccc}
\tabletypesize{\scriptsize}
\tablecaption{Previous Observations of HIP~79797~Bb with respect to HIP~79797~Ba}
\tablewidth{0pt}
\tablehead{
\colhead{Epoch} & \colhead{$\rho$ (")} & \colhead{$\theta$ ($^{\circ}$)} & \colhead{Reference}
}
\startdata
2010.27 & $0.060\,\pm0.020$ & $331\,\pm13$ & \citet{Nielsen:2013}\\
2010.35 & $0.060\,\pm0.006$ & $339\,\pm3$ & \citet{Nielsen:2013}\\
2011.35 & $0.063\,\pm0.015$ & $340\,\pm10$ & \citet{Nielsen:2013}\\
2012.26 & $0.063\,\pm0.007$ & $342\,\pm3$ & \citet{Nielsen:2013}\\
2012.66 & $0.077\,\pm0.014$ & $338\,\pm9$ & \citet{Nielsen:2013}\\
\hline
\\
& & Total Change in $\theta$: & 7\,$^{\circ}$ 
\label{tab:hip79797}
\end{deluxetable*}

\begin{deluxetable*}{cccc}
\tabletypesize{\scriptsize}
\tablecaption{Previous Observations of HR~3549~B}
\tablewidth{0pt}
\tablehead{
\colhead{Epoch} & \colhead{$\Delta\alpha$ (")} & \colhead{$\Delta$dec (")} & \colhead{Reference}
}
\startdata
2013.03 & $0.333\,\pm0.009$ & $−0.806\,\pm0.009$ & \citet{Mawet:2015}\\
2015.01 & $0.334\,\pm0.015$ & $−0.788\,\pm0.015$ & \citet{Mawet:2015}\\
2015.96 & $0.348\,\pm0.004$ & $-0.776\,\pm0.004$ & \citet{Mesa:2016}\\
2015.99 & $0.344\,\pm0.007$ & $-0.775\,\pm0.007$ & \citet{Mesa:2016}\\
\hline \\
& & Total Change in $\theta$: & 1.49\,$^{\circ}$ 
\label{tab:hr3549}
\end{deluxetable*}

\begin{deluxetable*}{cccc}
\tabletypesize{\scriptsize}
\tablecaption{Previous Observations of 2M1207 b}
\tablewidth{0pt}
\tablehead{
\colhead{Epoch} & \colhead{$\rho$ (")} & \colhead{$\theta$ ($^{\circ}$)} & \colhead{Reference}
}
\startdata
2004.32 & $0.772\,\pm0.004$ & $125.4\,\pm0.3$ & \citet{Chauvin:2005}\\
2004.66 & $0.7737\,\pm0.0022$ & $125.37\,\pm0.03$ & \citet{Song:2006}\\
2005.10 & $0.768\,\pm0.005$ & $125.4\,\pm0.3$ & \citet{Chauvin:2005}\\
2005.23 & $0.769\,\pm0.010$ & $125.6\,\pm0.7$ & \citet{Mohanty:2007}\\
2005.24 & $0.776\,\pm0.008$ & $125.5\,\pm0.3$ & \citet{Chauvin:2005}\\
2005.32 & $0.7735\,\pm0.0023$ & $125.61\,\pm0.20$ & \citet{Song:2006}\\
\hline \\
& & Total Change in $\theta$: & 0.21\,$^{\circ}$ 
\label{tab:2m1207}
\end{deluxetable*}

\begin{deluxetable*}{cccc}
\tabletypesize{\scriptsize}
\tablecaption{Previous Observations of $\kappa$~And~B}
\tablewidth{0pt}
\tablehead{
\colhead{Epoch} & \colhead{$\rho$ (")} & \colhead{$\theta$ ($^{\circ}$)} & \colhead{Reference}
}
\startdata
2012.00 & $1.070\,\pm0.010$ & $55.7\,\pm0.6$ & \citet{Carson:2013}\\
2012.52 & $1.058\,\pm0.007$ & $56.0\,\pm0.4$ & \citet{Carson:2013}\\
2012.83 & $1.029\,\pm0.005$ & $55.3\,\pm0.3$ & \citet{Bonnefoy:2014}$^a$\\
\hline \\
& & Total Change in $\theta$: & 0.4\,$^{\circ}$ 
\tablenotetext{a}{The measurements from Bonnefoy et al. were extracted from two nights of data. The date presented above is the average of the two observation dates.}
\label{tab:kand}
\end{deluxetable*}

\begin{deluxetable*}{cccc}
\tabletypesize{\scriptsize}
\tablecaption{Previous Observations of $\eta$~Tel~B}
\tablewidth{0pt}
\tablehead{
\colhead{Epoch} & \colhead{$\rho$ (")} & \colhead{$\theta$ ($^{\circ}$)} & \colhead{Reference}
}
\startdata
1998.49 & $4.170\,\pm 0.033$ & $166.95\,\pm 0.36$ & \citet{Lowrance:2000}; \citet{Neuhauser:2011}\\
2000.31 & $4.107\,\pm0.057$ & $166.90 \,\pm 0.42$  & \citet{Guenther:2001}; \citet{Neuhauser:2011}\\
2000.38 & $4.310\,\pm0.270$ & $165.8 \,\pm 6.7$  & \citet{Guenther:2001}\\
2004.33 & $4.189\,\pm0.020$ & $167.32 \,\pm 0.22$  & \citet{Neuhauser:2011}\\
2004.33 & $4.200\,\pm0.017$ & $166.85 \,\pm 0.22$  & \citet{Neuhauser:2011}\\
2004.34 & $4.199\,\pm0.036$ & $167.02 \,\pm 0.22$  & \citet{Neuhauser:2011}\\
2004.34 & $4.195\,\pm0.017$ & $166.97 \,\pm 0.22$  & \citet{Neuhauser:2011}\\
2006.43 & $4.170\,\pm0.110$ & $167.2 \,\pm 1.4$  & \citet{Geissler:2008}\\
2007.75 & $4.212\,\pm0.033$ & $167.42 \,\pm 0.35$  & \citet{Neuhauser:2011}\\
2008.31 & $4.214\,\pm0.017$ & $166.81 \,\pm 0.22$  & \citet{Neuhauser:2011}\\
2008.60 & $4.195\,\pm0.017$ & $166.87 \,\pm 0.29$  & \citet{Neuhauser:2011}\\
2008.60 & $4.194\,\pm0.016$ & $166.20 \,\pm 0.29$  & \citet{Neuhauser:2011}\\
2009.27 & $4.175\,\pm 0.09$ & $167.6\,\pm 0.2$ & \citet{Nielsen:2013}\\
2009.35 & $4.239\,\pm0.104$ & $168.5 \,\pm 1.3$  & \citet{Neuhauser:2011}\\
2009.50 & $4.199\,\pm0.031$ & $166.99 \,\pm 0.230$  & \citet{Neuhauser:2011}\\
\hline \\
& & Total Change in $\theta$: & 1.2\,$^{\circ}$
\label{tab:etaTel}
\end{deluxetable*}

\begin{deluxetable*}{cccc}
\tabletypesize{\scriptsize}
\tablecaption{Previous Observations of 2M~0103-55~(AB)~b}
\tablewidth{0pt}
\tablehead{
\colhead{Epoch} & \colhead{$\rho$ (")} & \colhead{$\theta$ ($^{\circ}$)} & \colhead{Reference}
}
\startdata
2002.82 & $1.718\,\pm0.015$ & $338.0\,\pm0.05$ & \citet{Delorme:2013}\\
2012.90 & $1.770\,\pm0.003$ & $336.1\,\pm0.01$ & \citet{Delorme:2013}\\
\hline \\
& & Total Change in $\theta$: & 1.9\,$^{\circ}$
\label{tab:2M0103}
\tablenotetext{}{2M~0103-55~(AB) is a binary, resolved in all epochs of astrometry. Astrometry above is measured relative to the barycenter of 2M~0103-55~(AB).}
\end{deluxetable*}

\begin{deluxetable*}{cccc}
\tabletypesize{\scriptsize}
\tablecaption{Previous Observations of CD-35~2722~B}
\tablewidth{0pt}
\tablehead{
\colhead{Epoch} & \colhead{$\rho$ (")} & \colhead{$\theta$ ($^{\circ}$)} & \colhead{Reference}
}
\startdata
2009.01 & $3.172\,\pm0.005$ & $244.13\,\pm0.25$ & \citet{Wahhaj:2011}\\
2010.02 & $3.137\,\pm0.005$ & $243.10\,\pm0.25$ & \citet{Wahhaj:2011}\\
\hline \\
& & Total Change in $\theta$: & 1.03\,$^{\circ}$ 
\label{tab:CD35}
\end{deluxetable*}

\begin{deluxetable*}{cccc}
\tabletypesize{\scriptsize}
\tablecaption{Previous Observations of GJ~504~b}
\tablewidth{0pt}
\tablehead{
\colhead{Epoch} & \colhead{$\rho$ (")} & \colhead{$\theta$ ($^{\circ}$)} & \colhead{Reference}
}
\startdata
2011.23 & $2.479\,\pm0.016$ & $327.94\,\pm0.39$ & \citet{Kuzuhara:2013}\\
2011.39 & $2.483\,\pm0.008$ & $327.45\,\pm0.19$ & \citet{Kuzuhara:2013}\\
2011.61 & $2.481\,\pm0.033$ & $326.84\,\pm0.94$ & \citet{Kuzuhara:2013}\\
2011.62 & $2.448\,\pm0.024$ & $325.82\,\pm0.66$ & \citet{Kuzuhara:2013}\\
2012.16 & $2.483\,\pm0.015$ & $326.46\,\pm0.36$ & \citet{Kuzuhara:2013}\\
2012.28 & $2.487\,\pm0.008$ & $326.54\,\pm0.18$ & \citet{Kuzuhara:2013}\\
2012.39 & $2.499\,\pm0.026$ & $326.14\,\pm0.61$ & \citet{Kuzuhara:2013}\\
\hline \\
& & Total Change in $\theta$: & 1.8\,$^{\circ}$
\label{tab:GJ504}
\end{deluxetable*}

\newpage

\begin{deluxetable*}{ccccccc}
\tabletypesize{\scriptsize}
\tablecaption{Orbit of HD~1160~B with respect to HD~1160~A}
\tablewidth{0pt}
\tablehead{
\colhead{orbital element} & \colhead{unit} &\colhead{max probability} & \colhead{min $\chi ^2$} & \colhead{median} & \colhead{68\,\% confidence range} & \colhead{95\,\% confidence range}
}
\startdata
a & au &45.27& 77.86& 77& 50-173& 42-834\\
P & yr&218.02& 482.21& 479& 252-1627& 194-17134\\
e & &0.8119& 0.2550& 0.77& 0.35-0.94& 0.05-0.98\\
i& $^{\circ}$ &109.3& 98.2& 103& 96-119& 92-149\\
$\omega$ & $^{\circ}$ &10.2& 99.9& 96& 39-143& 6-174\\
$\Omega$ & $^{\circ}$ &66.1& 61.1& 63& 38-91& 9-169\\
$T_0$ & yr &2137.84& 2138.27& 2220.06& 2131.28-2796.96& 2087.36-9871.18\\
\label{tab:hd1160b_outputs}
\tablenotetext{}{The acceptance rate was 0.24\,\%.}
\end{deluxetable*}

\begin{figure}
    \centering
    \includegraphics[width=\textwidth]{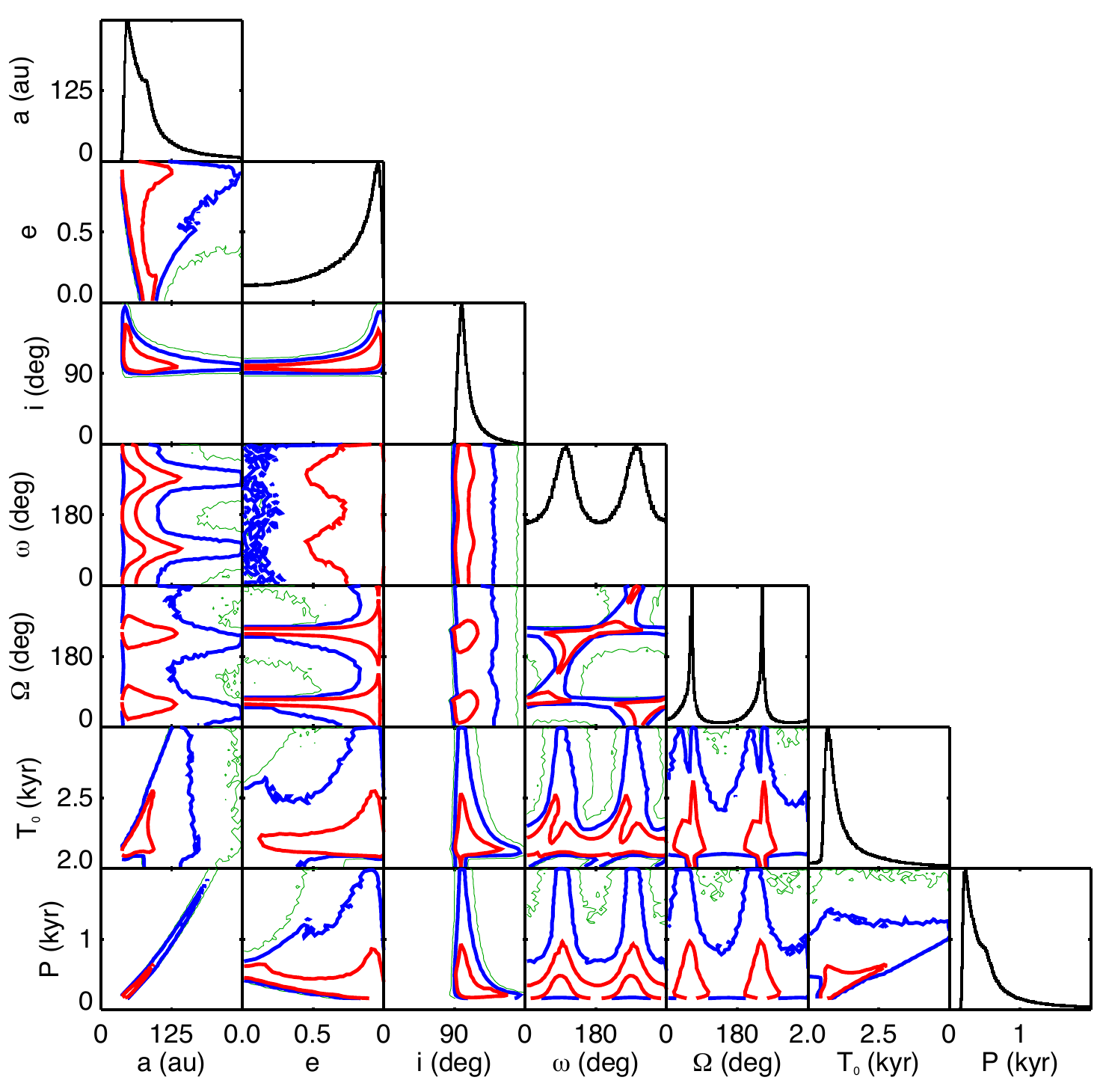}
    \caption{Triangle plots for the orbit of HD~1160~B with respect to HD~1160~A. See Figure \ref{fig:gj504_covariance}.}
    \label{fig:hd1160b_covariance}
\end{figure}

\begin{figure}
\centering
\includegraphics[width=\textwidth]{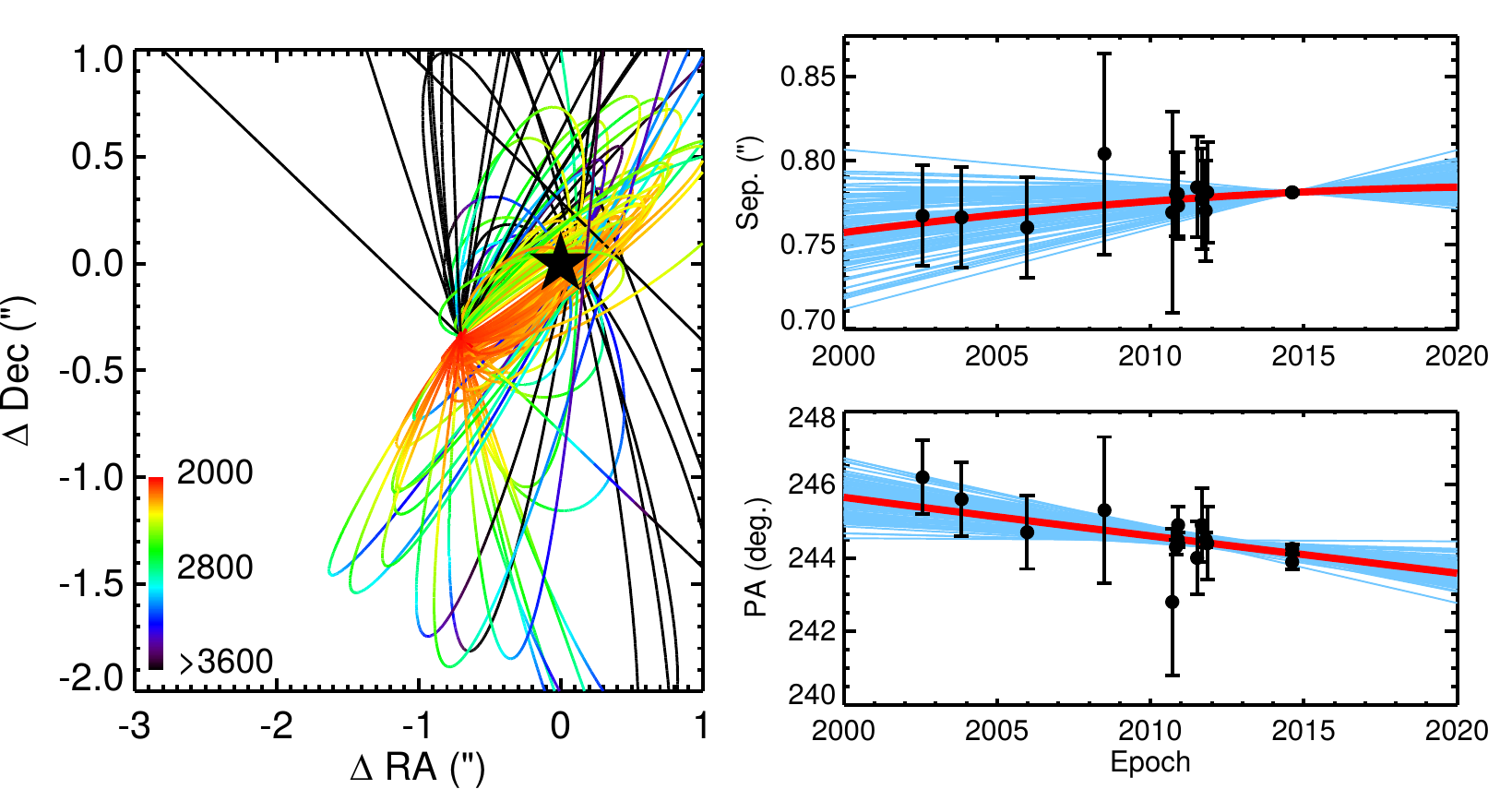}
\caption{Depictions of 100 likely orbits fit to relative astrometry of HD~1160~B with repsect to HD~1160~A. See Figure \ref{fig:gj504_orbit}.}
\label{fig:hd1160b_orbit}
\end{figure}

\begin{deluxetable*}{ccccccc}
\tabletypesize{\scriptsize}
\tablecaption{Orbit of HD~1160~C with respect to HD~1160~A}
\tablewidth{0pt}
\tablehead{
\colhead{orbital element} & \colhead{unit} &\colhead{max probability} & \colhead{min $\chi ^2$} & \colhead{median} & \colhead{68\,\% confidence range} & \colhead{95\,\% confidence range}
}
\startdata
a & au&1631.43& 2512.52& 651& 491-1083& 372-2454\\
P & yr &45119.44& 80174.59& 11260& 7362-24190& 4852-82443\\
e & &0.6286& 0.8110& 0.33& 0.11-0.60& 0.02-0.82\\
i & $^{\circ}$ &136.1& 170.9& 146& 128-163& 110-174\\
$\omega$  & $^{\circ}$&31.4& 93.1& 84& 28-150& 4-176\\
$\Omega$ & $^{\circ}$&25.2& 56.4& 62& 20-148& 3-177\\
$T_0$ & yr &46827.45& 2374.57& 3812.12& 2915.07-8680.52& 2195.37-31708.72\\
\label{tab:hd1160c_outputs}
\tablenotetext{}{Note: The acceptance rate was 0.006\,\%.}
\end{deluxetable*}

\begin{figure}
\includegraphics[width=\textwidth]{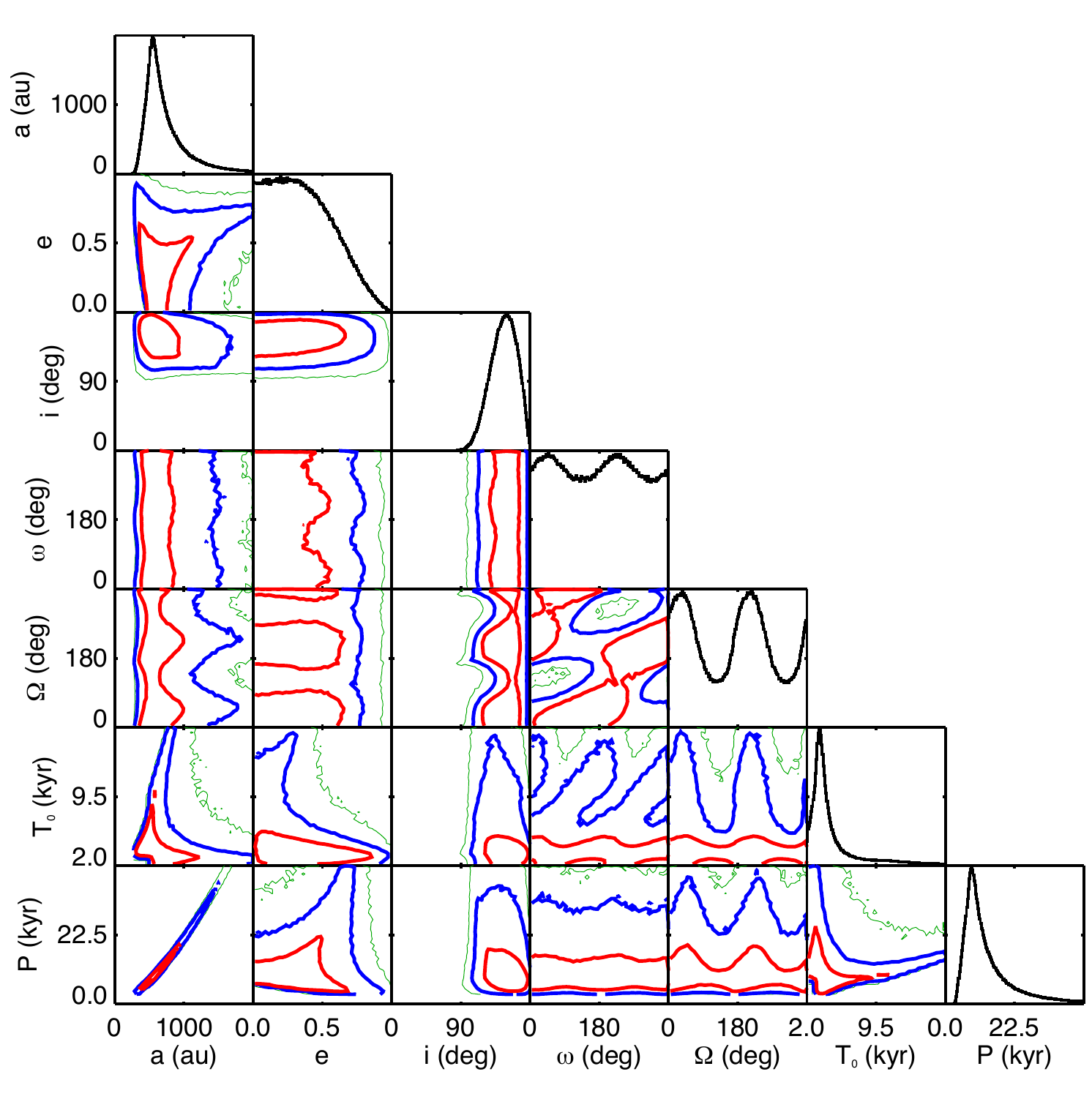}
\caption{Triangle plots for the orbit of HD~1160~C with respect to HD~1160~A. See Figure \ref{fig:gj504_covariance}.}
\label{fig:hd1160c_covariance}
\end{figure}

\begin{figure}
\includegraphics[width=\textwidth]{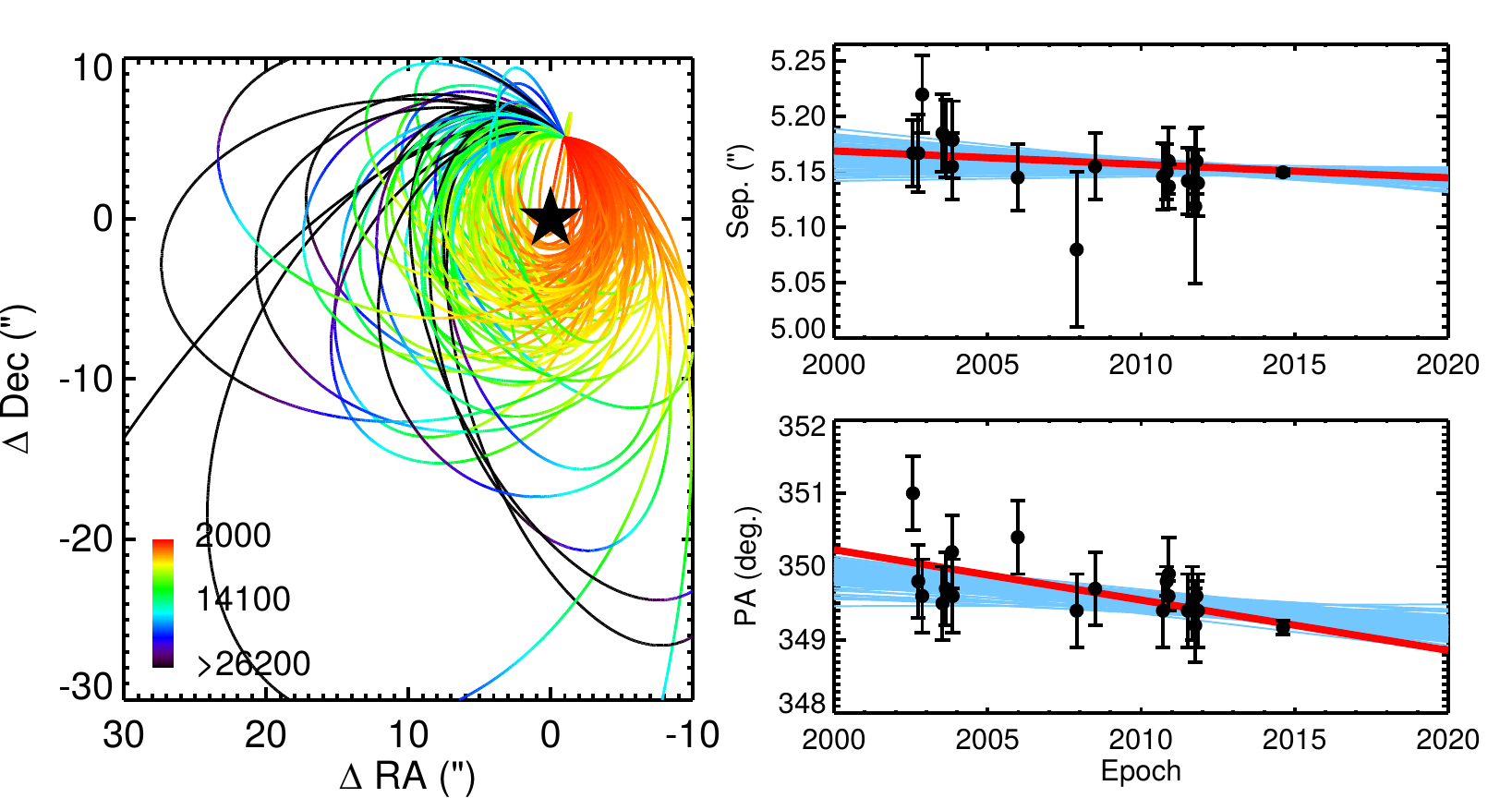}
\caption{Depictions of 100 likely orbits fit to relative astrometry of HD~1160~C with respect to HD~1160~A. See Figure \ref{fig:gj504_orbit}.}
\label{fig:hd1160c_orbit}
\end{figure}

\begin{deluxetable*}{ccccccc}
\tabletypesize{\scriptsize}
\tablecaption{Orbit of HIP~79797~Bb with respect to HIP~79797~Ba}
\tablewidth{0pt}
\tablehead{
\colhead{orbital element} & \colhead{unit} & \colhead{max probability} & \colhead{min $\chi ^2$} & \colhead{median} & \colhead{68\,\% confidence range} & \colhead{95\,\% confidence range}
}
\startdata
$a$ & au &2.01& 9.96& 3& 2-8& 2-35\\
$P$ & yr &8.71& 82.70& 23& 11-86& 7-735\\
$e$ & &0.9133& 0.8965& 0.75& 0.31-0.95& 0.05-0.99\\
$i$ &$^{\circ}$ &75.2& 85.0& 83& 66-93& 34-117\\
$\omega$ & $^{\circ}$ &151.1& 104.0& 95& 39-145& 6-174\\
$\Omega$ &$^{\circ}$&168.4& 8.3& 157& 51-166& 4-177\\
$T_0$ & yr &1998.44& 2061.23& 2005.47& 1998.34-2034.01& 1995.48-2338.70\\
\label{tab:79797_outputs}
\tablenotetext{}{Note: The acceptance rate was 13.8\,\%.}
\end{deluxetable*}

\begin{figure}
\includegraphics[width=\textwidth]{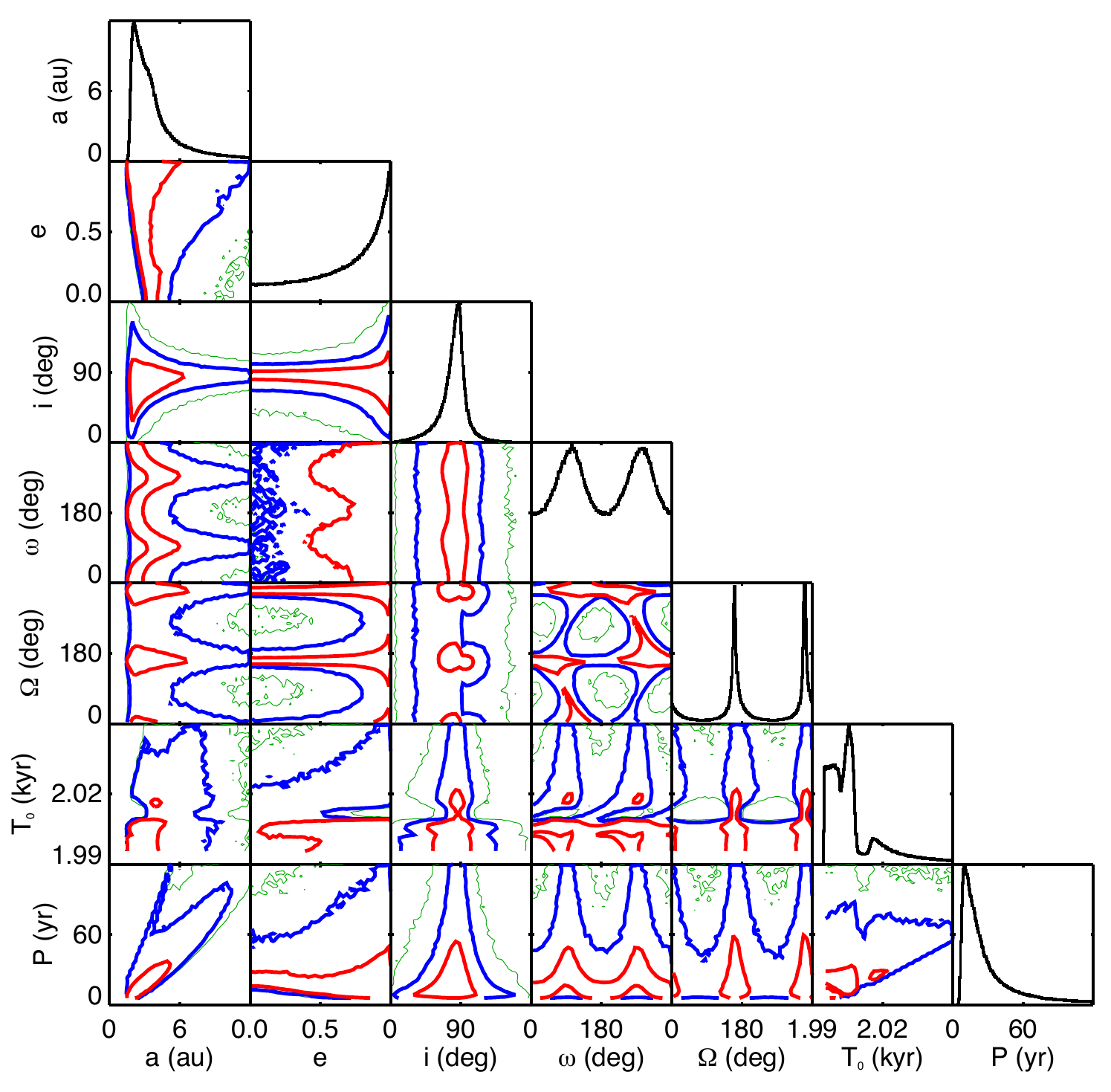}
\caption{Triangle plots for the orbit of HIP~79797~Bb with respect to HIP~79797~Ba. See Figure \ref{fig:gj504_covariance}.}
\label{fig:hip79797_covplot}
\end{figure}

\begin{figure}
\includegraphics[width=\textwidth]{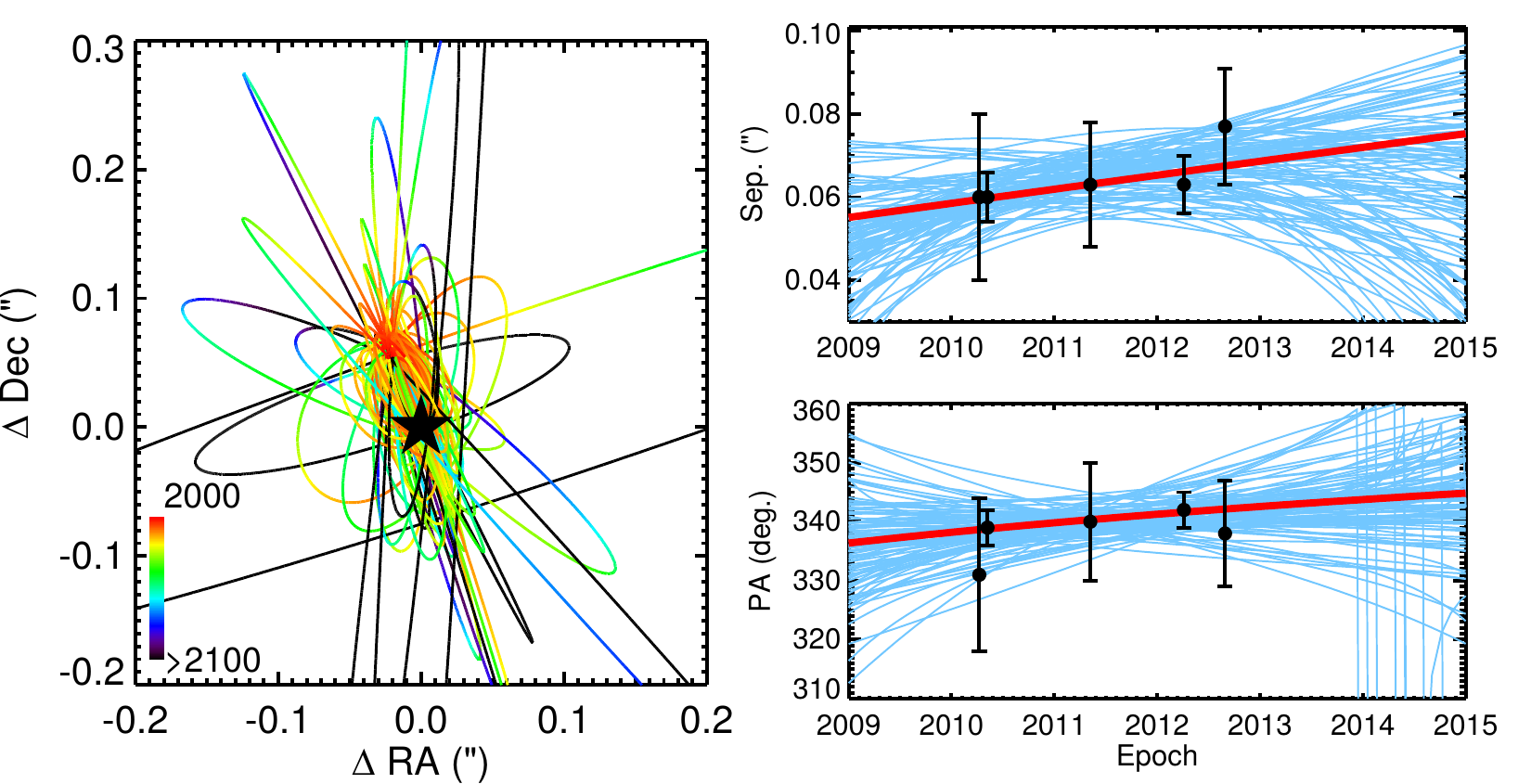}
\caption{Depictions of 100 likely orbits fit to relative astrometry of HIP~79797~Bb with respect to HIP~79797~Ba. See figure \ref{fig:gj504_orbit}. Note: the vertical bars in the bottom right plot result from PA crossing 360\,$^{\circ}$.}
\label{fig:hip79797_orbit}
\end{figure}

\begin{deluxetable*}{ccccccc}
\tabletypesize{\scriptsize}
\tablecaption{Orbit of HR~3549~B with respect to HR~3549~A}
\tablewidth{0pt}
\tablehead{
\colhead{orbital element} & \colhead{unit} & \colhead{max probability} & \colhead{min $\chi ^2$} & \colhead{median} & \colhead{68\,\% confidence range} & \colhead{95\,\% confidence range}
}
\startdata
$a$ & au &85.84& 83.73& 94& 66-159& 50-360\\
$P$ & yr &520.01& 496.86& 592& 346-1303& 231-4461\\
$e$ & &0.5103& 0.4797& 0.47& 0.16-0.75& 0.02-0.92\\
$i$ & $^{\circ}$ &138.5& 138.1& 130& 111-152& 96-170\\
$\omega$ & $^{\circ}$ &136.8& 167.3& 87& 26-152& 4-176\\
$\Omega$ & $^{\circ}$ &172.6& 14.0& 77& 13-168& 2-178\\
$T_0$ & yr &2094.84& 2105.86& 2115.38& 2078.47-2319.26& 2030.61-3633.32\\
\label{tab:hr3549_outputs}
\tablenotetext{}{Note: The acceptance rate was 2.59\,\%.}
\end{deluxetable*}

\begin{figure}
\includegraphics[width=\textwidth]{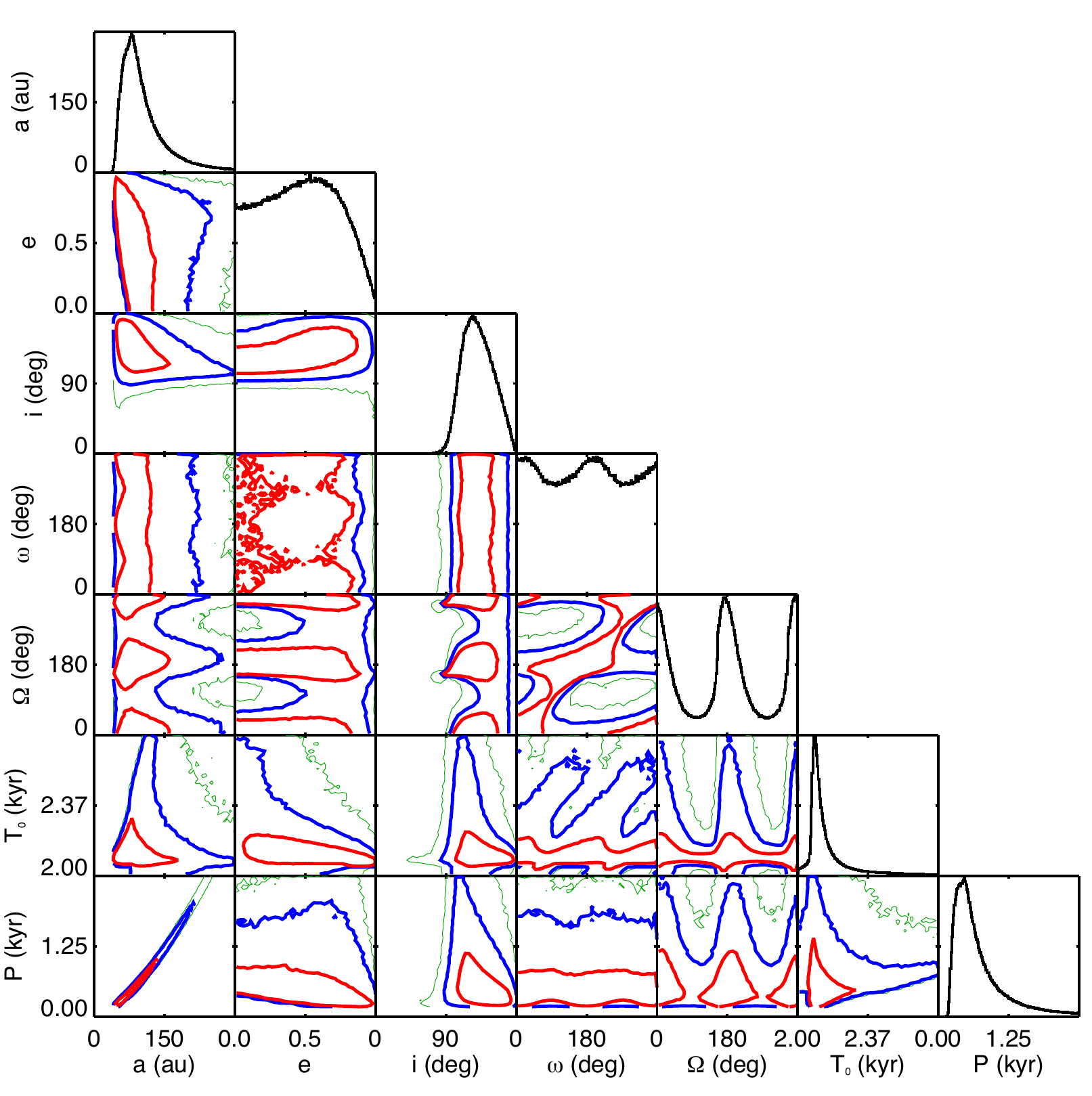}
\caption{Triangle plots for the orbit of HR~3549~B with respect to HR~3549~A. See Figure \ref{fig:gj504_covariance}.}
\label{fig:hr3549_covplot}
\end{figure}

\begin{figure}
\includegraphics[width=\textwidth]{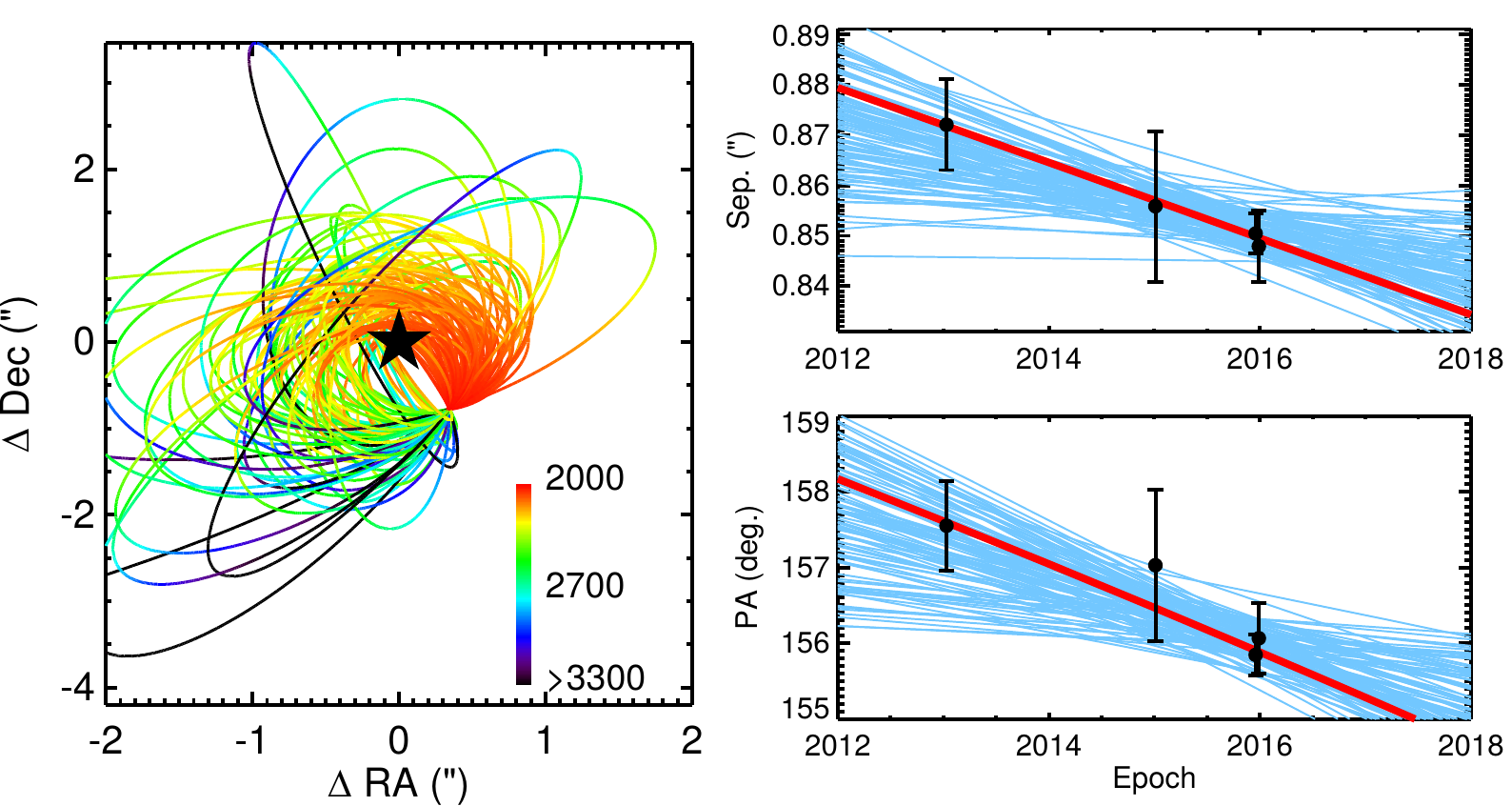}
\caption{Depictions of 100 likely orbits fit to relative astrometry of HR~3549~B with respect to HR~3549~A. See figure \ref{fig:gj504_orbit}.}
\label{fig:hr3549_orbit}
\end{figure}

\begin{deluxetable*}{ccccccc}
\tabletypesize{\scriptsize}
\tablecaption{Orbit of 2M~1207~b with respect to 2M~1207~A}
\tablewidth{0pt}
\tablehead{
\colhead{orbital element} & \colhead{unit} & \colhead{max probability} & \colhead{min $\chi ^2$} & \colhead{median} & \colhead{68\,\% confidence range} & \colhead{95\,\% confidence range}
}
\startdata
$a$ & au &35.26& 91.88& 46& 31-84& 24-231\\
$P$ & yr &1153.44& 4648.27& 1782& 974-4413& 633-20046\\
$e$ & &0.2226& 0.5226& 0.49& 0.15-0.83& 0.02-0.98\\
$i$ & $^{\circ}$ &41.6& 41.5& 69& 36-109& 13-150\\
$\omega$ & $^{\circ}$  &141.4& 161.6& 90& 29-151& 4-176\\
$\Omega$ & $^{\circ}$ &129.6& 1.9& 119& 52-146& 7-174\\
$T_0$ & yr &2424.39& 2172.58& 2683.34& 2285.03-4277.65& 2107.69-12883.36\\
\label{tab:2m1207_outputs}
\tablenotetext{}{Note: The acceptance rate was 13.99\,\%.}
\end{deluxetable*}

\begin{figure}
    \centering
    \includegraphics[width=\textwidth]{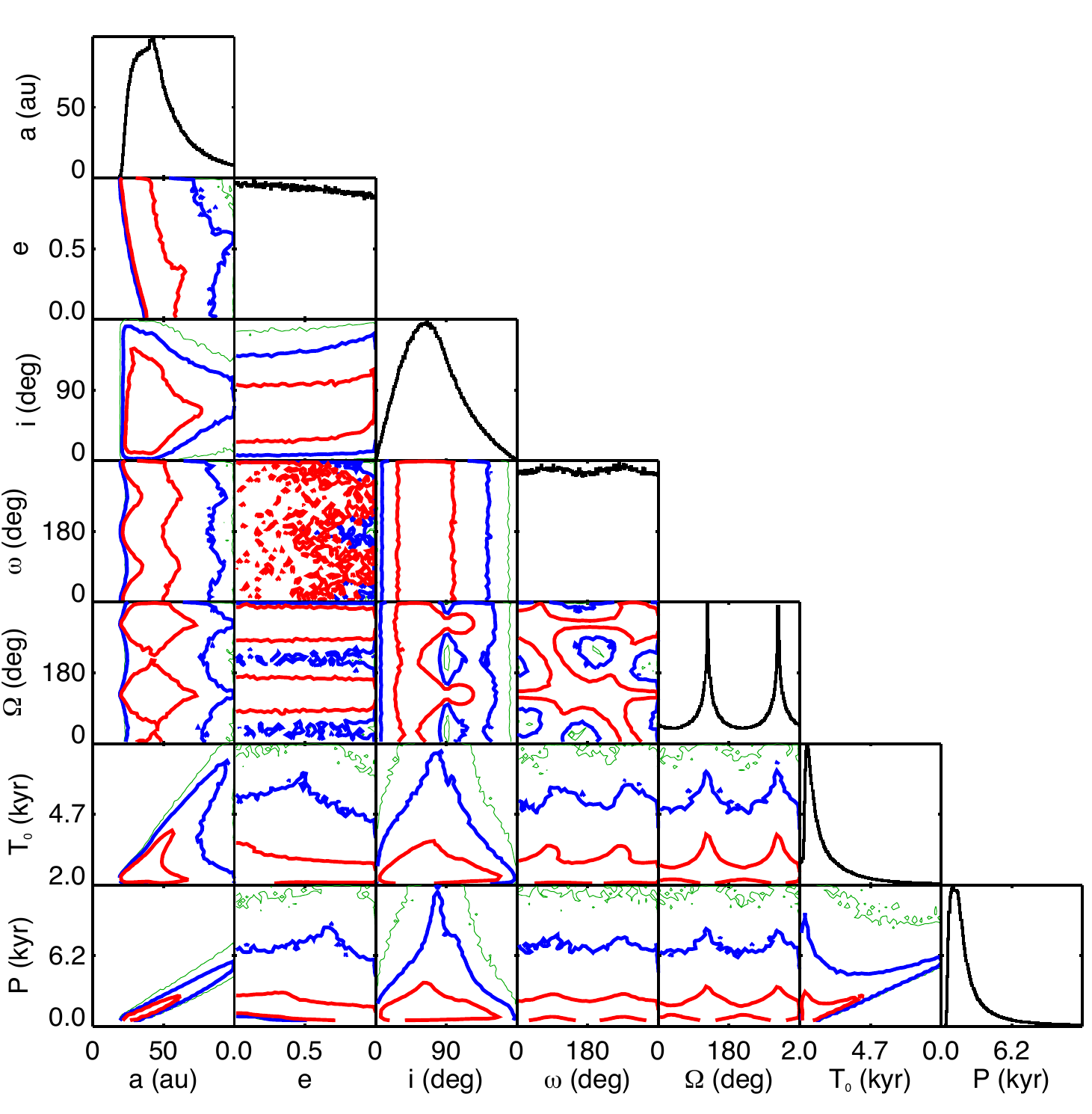}
    \caption{Triangle plots for the orbit of 2M~1207~b with respect to 2M~1207~A. See Figure \ref{fig:gj504_covariance}.}
    \label{fig:2m1207_covariance}
\end{figure}
\begin{figure}
    \includegraphics[width=\textwidth]{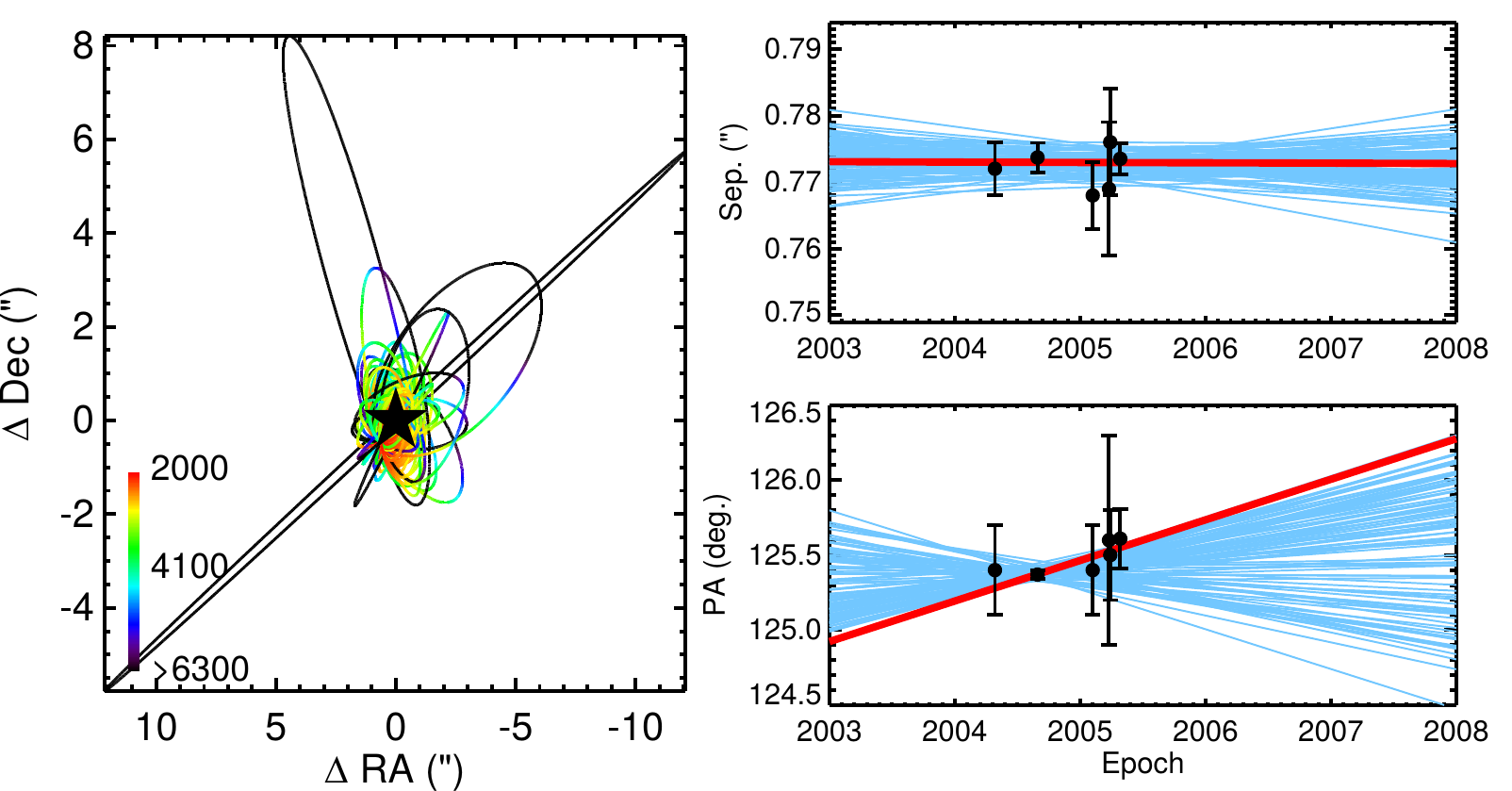}
    \caption{Depictions of 100 likely orbits fit to relative astrometry of 2M~1207~b with respect to 2M~1207~A.  See Figure \ref{fig:gj504_orbit}.}
    \label{fig:2m1207_orbit}
\end{figure}

\begin{deluxetable*}{ccccccc}
\tabletypesize{\scriptsize}
\tablecaption{Orbit of $\kappa$~And~B with respect to $\kappa$~And~B}
\tablewidth{0pt}
\tablehead{
\colhead{orbital element} & \colhead{unit} &\colhead{max probability} & \colhead{min $\chi ^2$} & \colhead{median} & \colhead{68\,\% confidence range} & \colhead{95\,\% confidence range}
}
\startdata
$a$ & au  &184.96& 184.96& 77& 54-123& 40-236\\
$P$ & yr &1385.97& 1385.97& 378& 223-768& 144-2033\\
$e$ & &0.8908& 0.8908& 0.41& 0.12-0.70& 0.02-0.85\\
$i$ & $^{\circ}$ &116.6& 116.6& 101& 83-125& 59-159\\
$\omega$ & $^{\circ}$ &138.7& 138.7& 112& 29-159& 3-177\\
$\Omega$ & $^{\circ}$ &67.8& 67.8& 63& 49-84& 26-127\\
$T_0$ & yr &2040.10& 2040.10& 2065.27& 2043.38-2188.31& 2015.35-2858.45\\
\label{tab:kapAnd_outputs}
\tablenotetext{}{Note: The acceptance rate was 0.02\,\%.}
\end{deluxetable*}

\begin{figure}
    \centering
    \includegraphics[width=\textwidth]{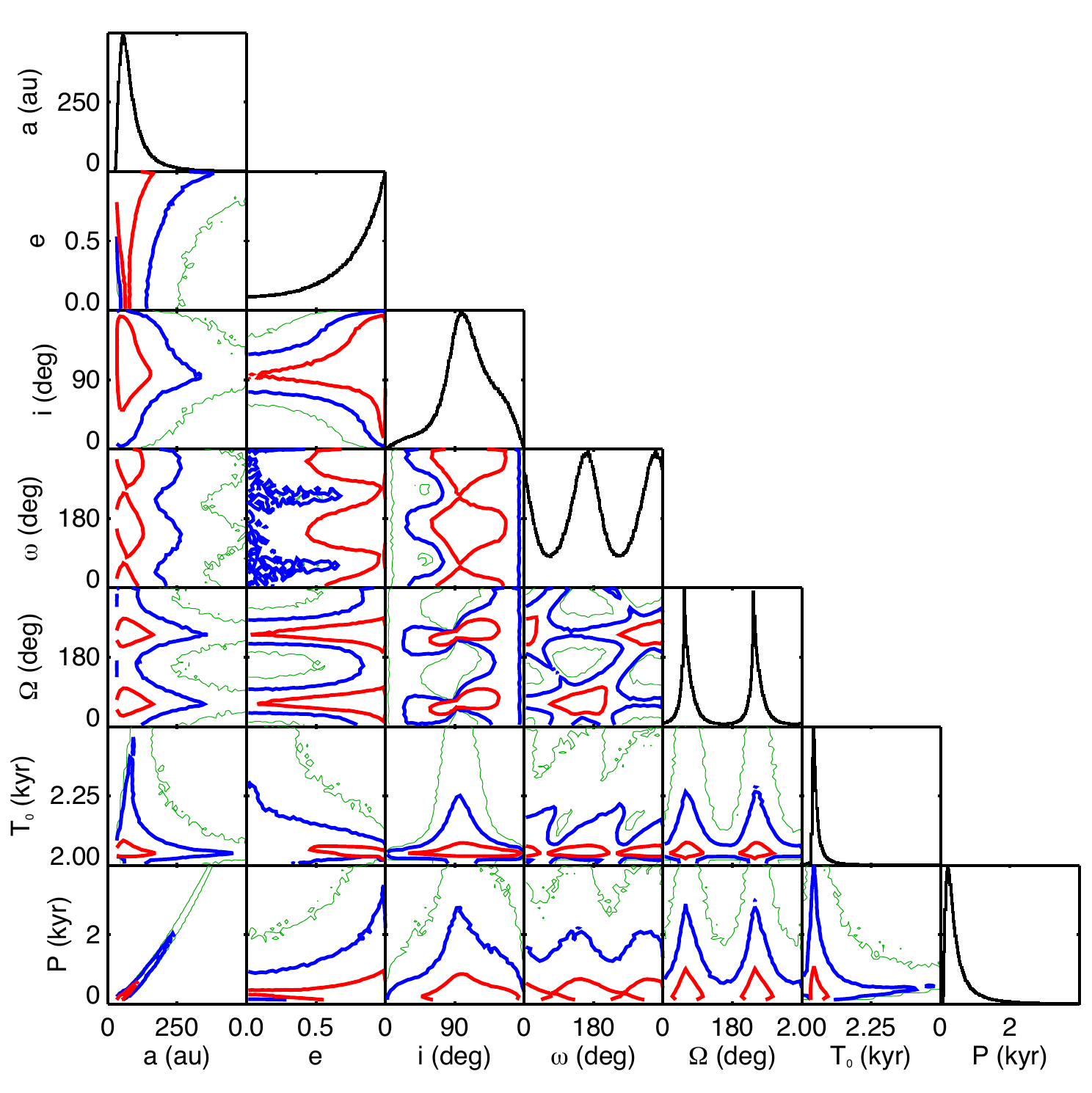}
    \caption{Triangle plots for the orbit of $\kappa$~And~B with respect to $\kappa$~And~A. See Figure \ref{fig:gj504_covariance}.}
    \label{fig:kapand_covariance}
\end{figure}
\begin{figure}
    \includegraphics[width=\textwidth]{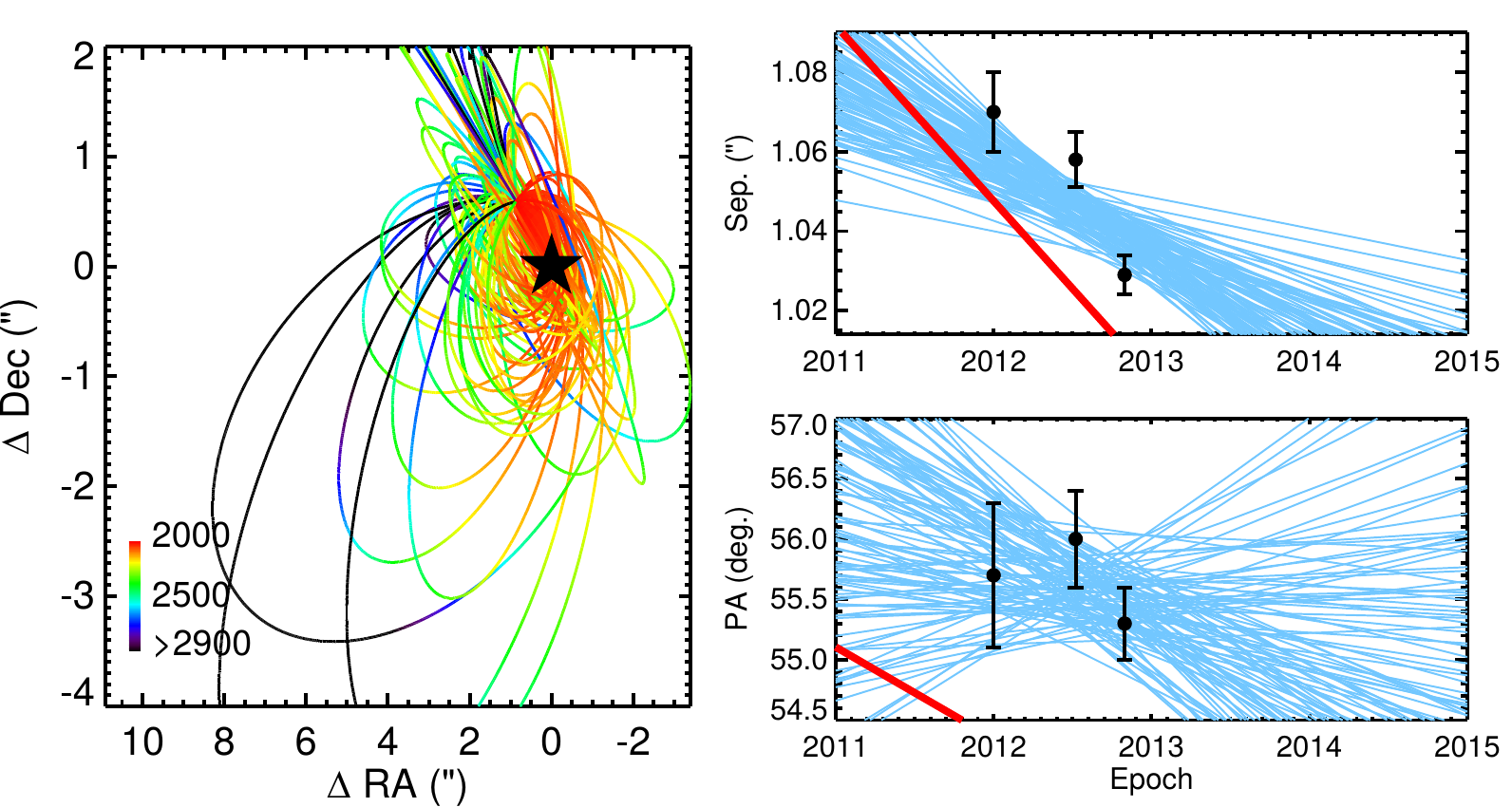}
    \caption{Depictions of 100 likely orbits fit to relative astrometry of $\kappa$~And~B with respect to $\kappa$~And~A. See figure \ref{fig:gj504_orbit}.}
    \label{fig:kapand_orbit}
\end{figure}

\begin{deluxetable*}{ccccccc}
\tabletypesize{\scriptsize}
\tablecaption{Orbit of $\eta$~Tel~B with respect to $\eta$~Tel~A}
\tablewidth{0pt}
\tablehead{
\colhead{orbital element} & \colhead{unit} & \colhead{max probability} & \colhead{min $\chi ^2$} & \colhead{median} & \colhead{68\,\% confidence range} & \colhead{95\,\% confidence range}
}
\startdata
$a$ & au &206.39& 207.86& 192& 125-432& 106-2091\\
$P$ & yr &1600.73& 1640.25& 1493& 781-5028& 612-53621\\
$e$ & &0.9084& 0.0354& 0.77& 0.34-0.96& 0.05-1.00\\
$i$ & $^{\circ}$ &88.4& 87.4& 86& 72-96& 40-120\\
$\omega$ & $^{\circ}$ &121.8& 165.6& 98& 37-146& 5-175\\
$\Omega$ & $^{\circ}$ &169.8& 167.4& 165& 42-170& 3-178\\
$T_0$ & yr &2851.39& 3623.16& 2669.33& 2391.73-4459.40& 2263.35-26828.14\\
\label{tab:etaTel_outputs}
\tablenotetext{}{Note: The acceptance rate was 0.00002\,\%.}
\end{deluxetable*}

\begin{figure}
    \centering
    \includegraphics[width=\textwidth]{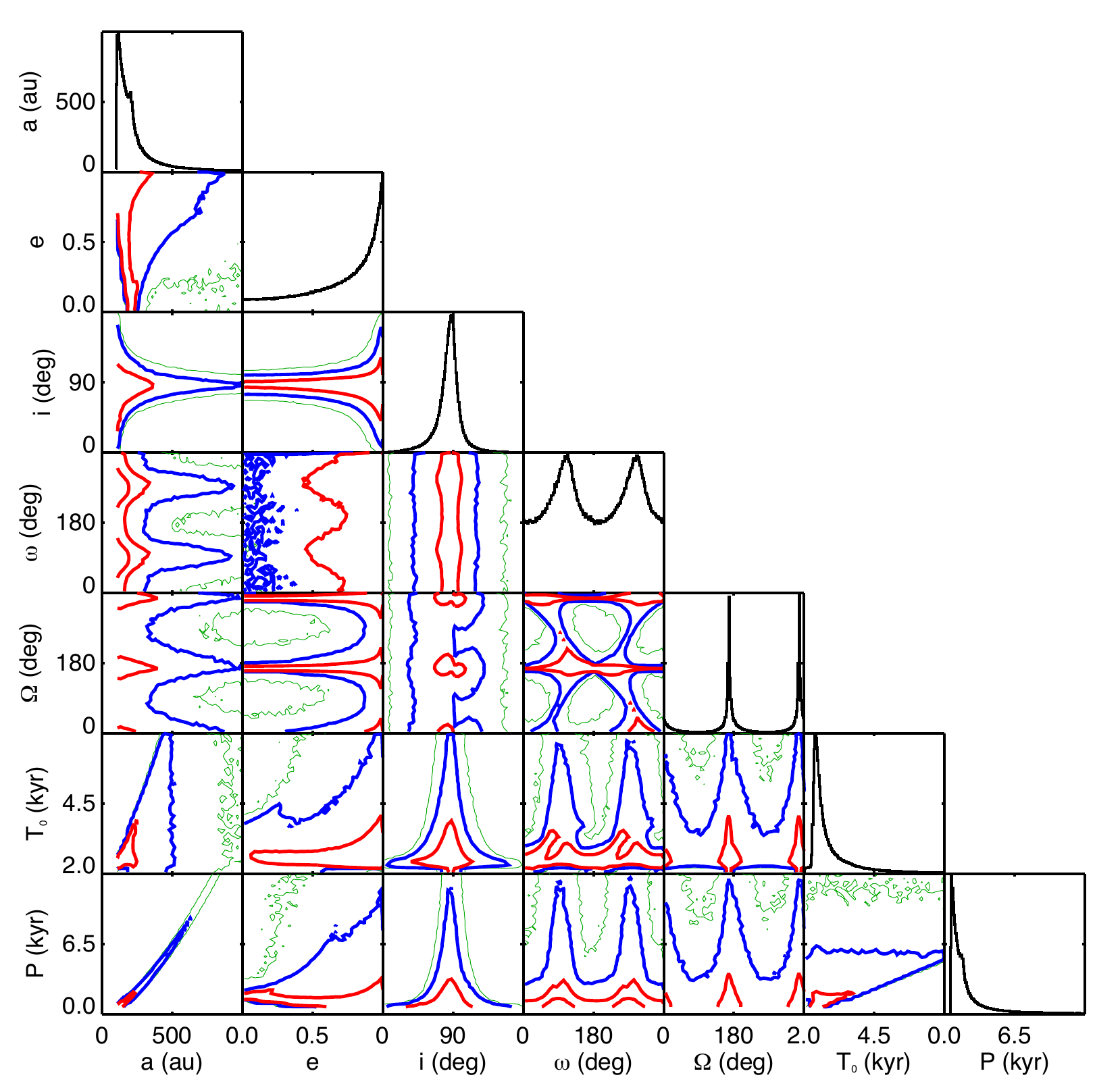}
    \caption{Triangle plots for the orbit of $\eta$~Tel~B with respect to $\eta$~Tel~A. See Figure \ref{fig:gj504_covariance}.}
    \label{fig:etatel_covariance}
\end{figure}
\begin{figure}
    \includegraphics[width=\textwidth]{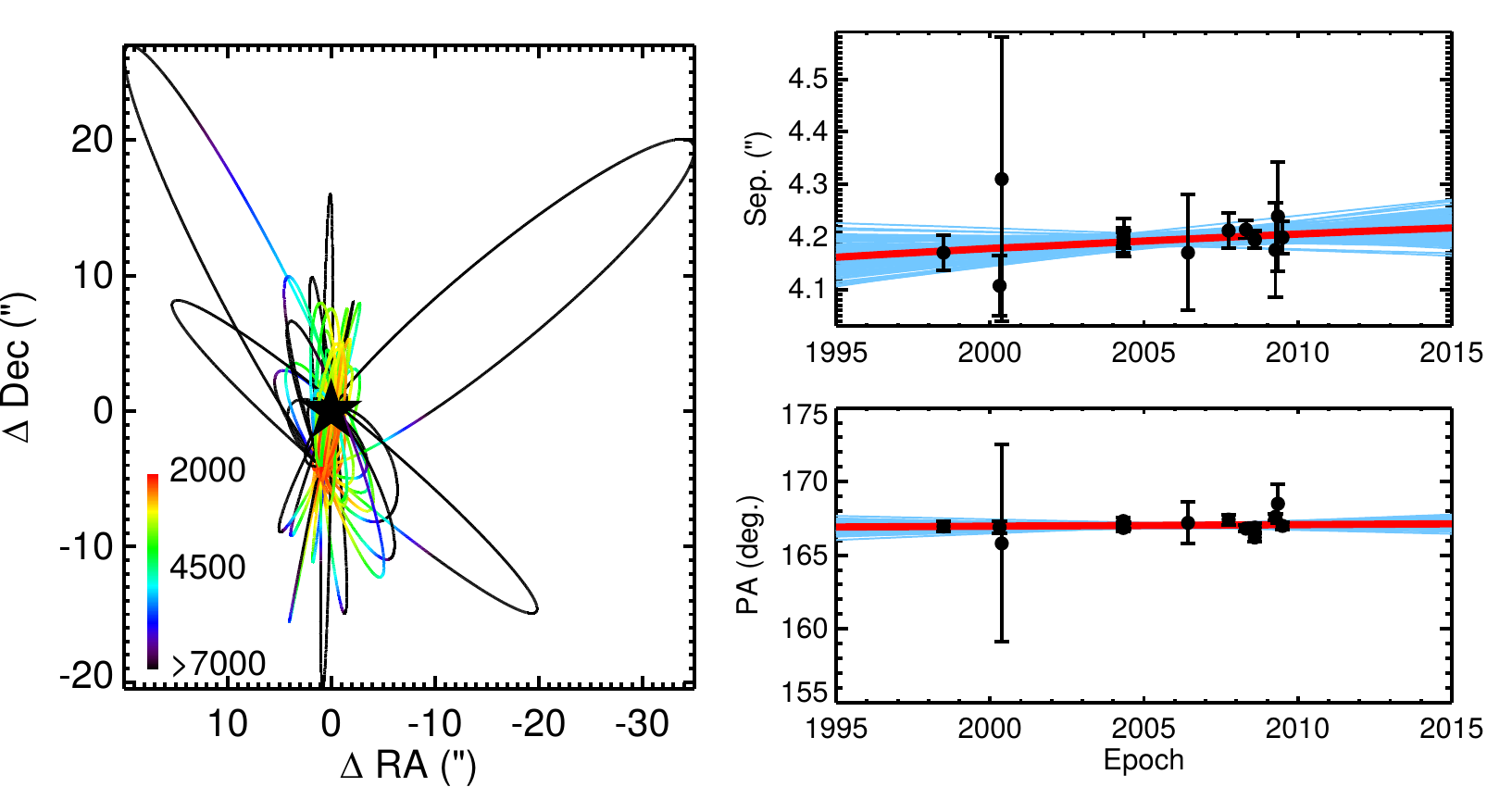}
    \caption{Depictions of 100 likely orbits fit to relative astrometry of $\eta$~Tel~B with respect to $\eta$~Tel~A. See figure \ref{fig:gj504_orbit}.}
    \label{fig:etaTel_orbit}
\end{figure}

\begin{deluxetable*}{ccccccc}
\tabletypesize{\scriptsize}
\tablecaption{Orbit of 2M~0103-55~(AB)~b with respect to 2M~0103-55~(AB)}
\tablewidth{0pt}
\tablehead{
\colhead{orbital element} & \colhead{unit} & \colhead{max probability} & \colhead{min $\chi ^2$} & \colhead{median} & \colhead{68\,\% confidence range} & \colhead{95\,\% confidence range}
}
\startdata
$a$ & au &104.92& 134.46& 102& 75-149& 58-256\\
$P$ & yr &1746.65& 2337.20& 1682& 1054-2990& 716-6723\\
$e$ & &0.1233& 0.2839& 0.32& 0.09-0.59& 0.01-0.74\\
$i$ & $^{\circ}$ &123.6& 115.6& 127& 119-144& 112-165\\
$\omega$ & $^{\circ}$ &34.3& 145.7& 87& 25-155& 3-177\\
$\Omega$ & $^{\circ}$ &124.4& 136.9& 122& 98-143& 22-167\\
$T_0$ & yr &3355.07& 2028.69& 3081.80& 2534.01-4267.75& 2068.56-7551.37\\
\label{tab:2m0103_outputs}
\tablenotetext{}{Note: The acceptance rate was 0.32\,\%.}
\end{deluxetable*}

\begin{figure}
    \centering
    \includegraphics[width=\textwidth]{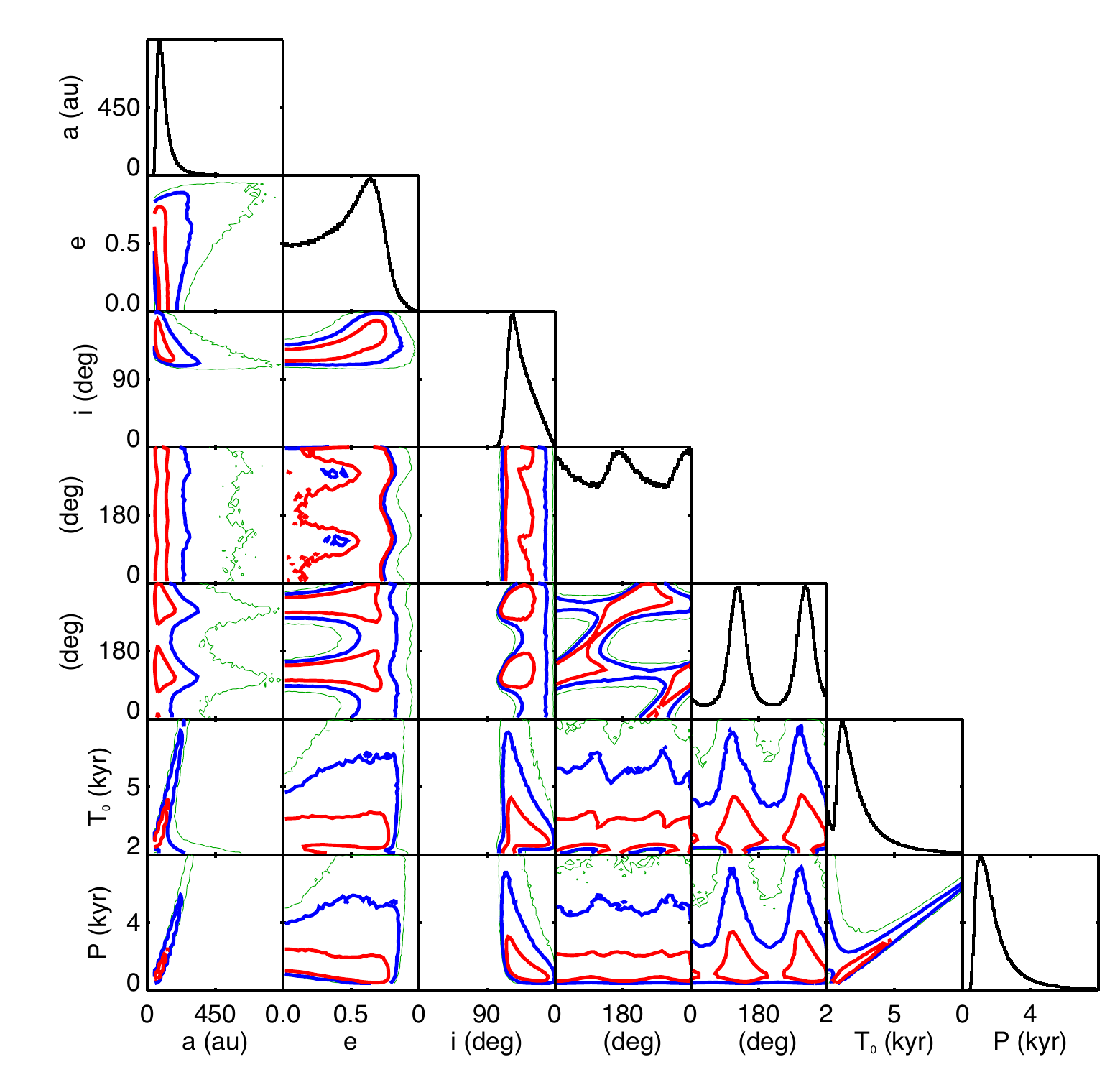}
    \caption{Triangle plots for the orbit of 2M~0103-55~(AB)~b with respect to 2M~0103-55~(AB). See Figure \ref{fig:gj504_covariance}.}
    \label{fig:2m0103_covariance}
\end{figure}
\begin{figure}
    \includegraphics[width=\textwidth]{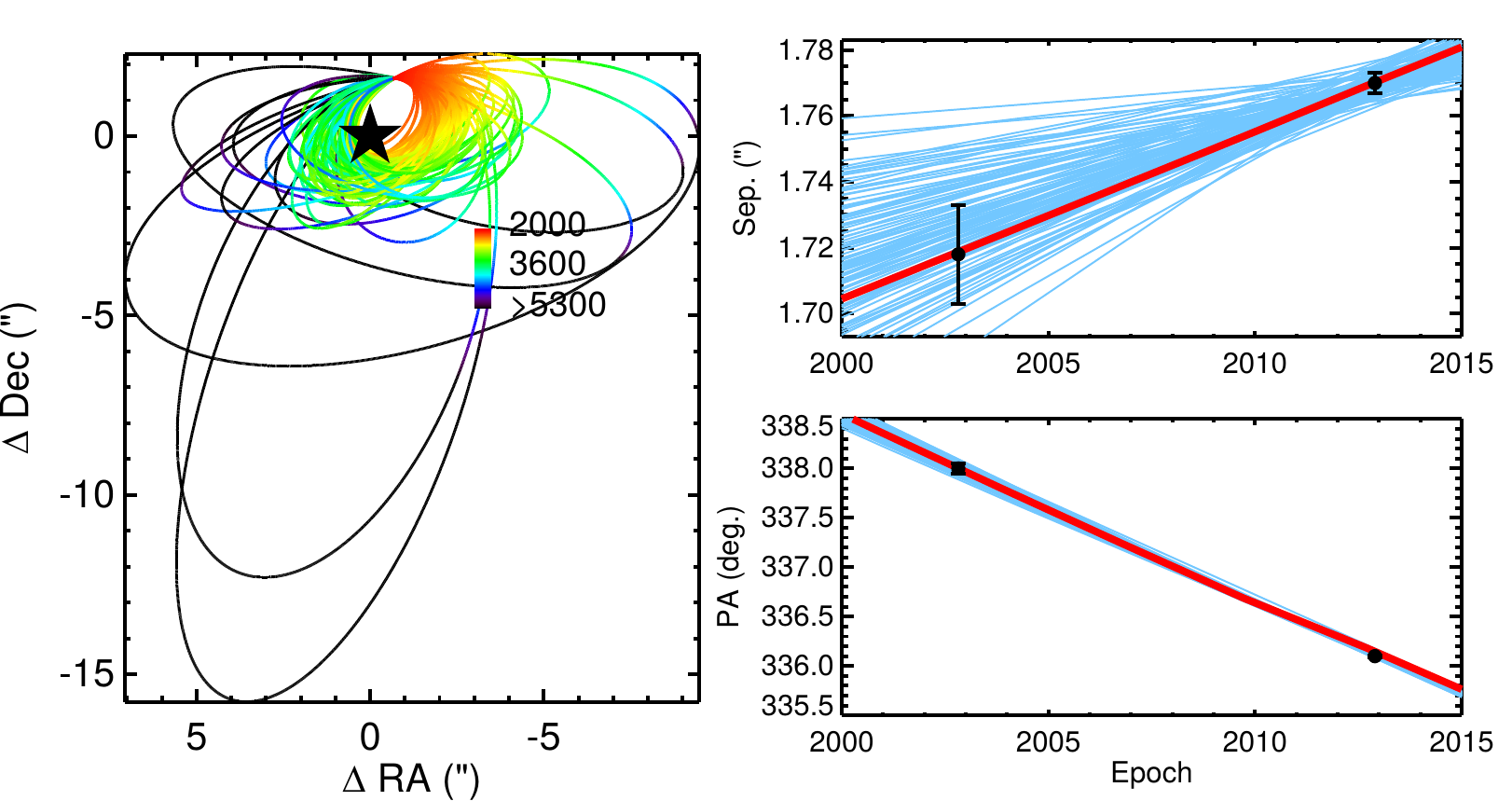}
    \caption{Depictions of 100 likely orbits fit to relative astrometry of 2M~0103-55~(AB)~b with respect to 2M~0103-55~(AB). See figure \ref{fig:gj504_orbit}.}
    \label{fig:2m0103_orbit}
\end{figure}

\begin{deluxetable*}{ccccccc}
\tabletypesize{\scriptsize}
\tablecaption{Orbit of CD-35~2722~B with respect to CD-35~2722~A}
\tablewidth{0pt}
\tablehead{
\colhead{orbital element} & \colhead{unit} & \colhead{max probability} & \colhead{min $\chi ^2$} & \colhead{median} & \colhead{68\,\% confidence range} & \colhead{95\,\% confidence range}
}
\startdata
$a$ & au &286.48& 10048.78& 115& 74-216& 53-520\\
$P$ & yr &6858.18& 1454272.75& 1853& 947-4772& 580-17796\\
$e$ & &0.9058& 0.9973& 0.82& 0.57-0.95& 0.18-0.99\\
$i$ & $^{\circ}$ &160.4& 154.2& 126& 95-156& 49-172\\
$\omega$ & $^{\circ}$ &110.4& 136.7& 128& 32-163& 3-177\\
$\Omega$ & $^{\circ}$ &72.1& 98.2& 75& 58-104& 20-158\\
$T_0$ & yr &2088.36& 2087.50& 2125.23& 2099.91-2184.95& 2082.76-2531.59\\
\label{tab:cd35_outputs}
\tablenotetext{}{Note: The acceptance rate was 0.006\,\%.}
\end{deluxetable*}

\begin{figure}
    \includegraphics[width=\textwidth]{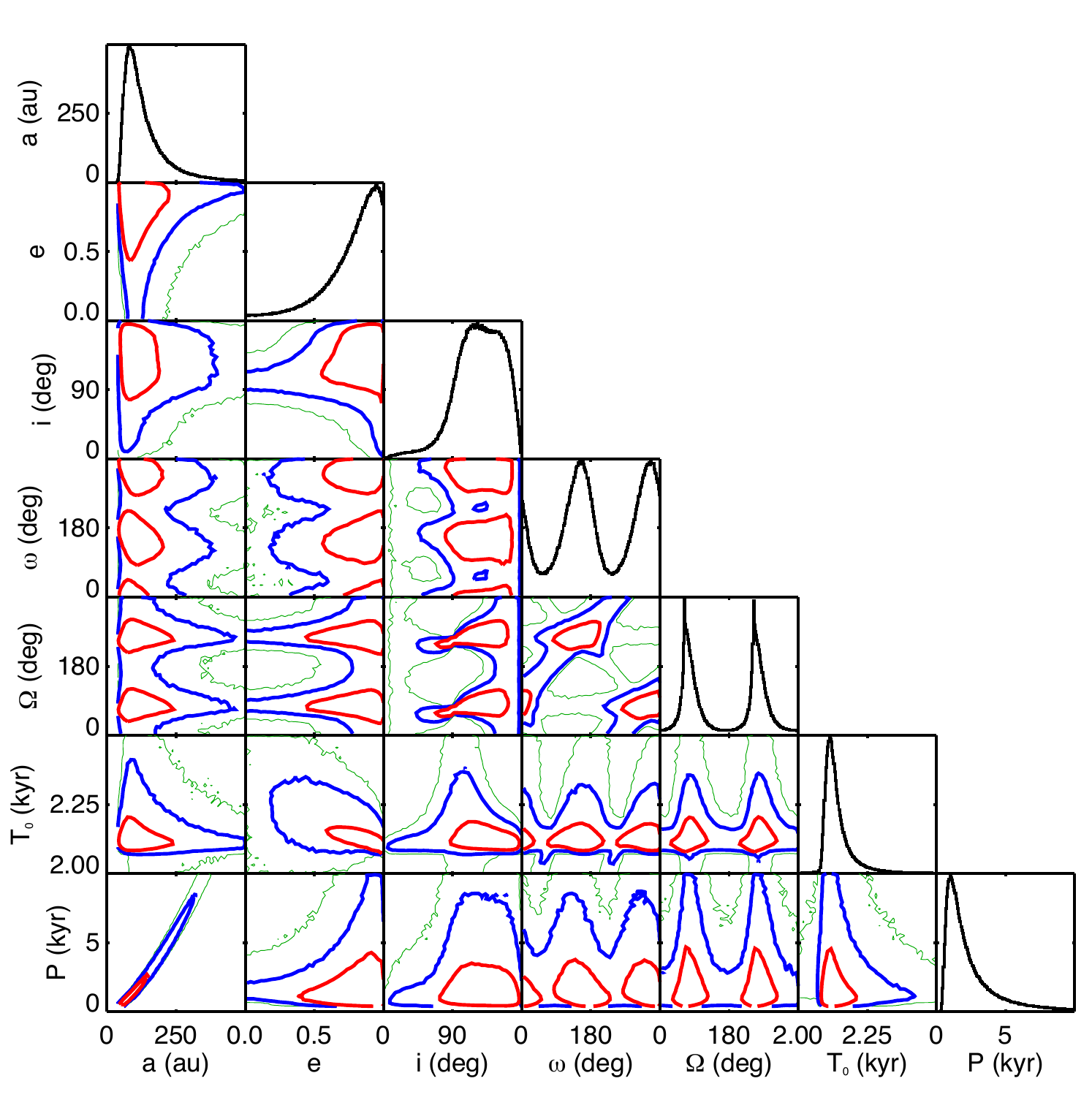}
    \caption{Triangle plots for the orbit of CD-35~2722~B with respect to CD-35~2722~A. See Figure \ref{fig:gj504_covariance}.}
    \label{fig:cd35_covariance}
\end{figure}

\begin{figure}
    \includegraphics[width=\textwidth]{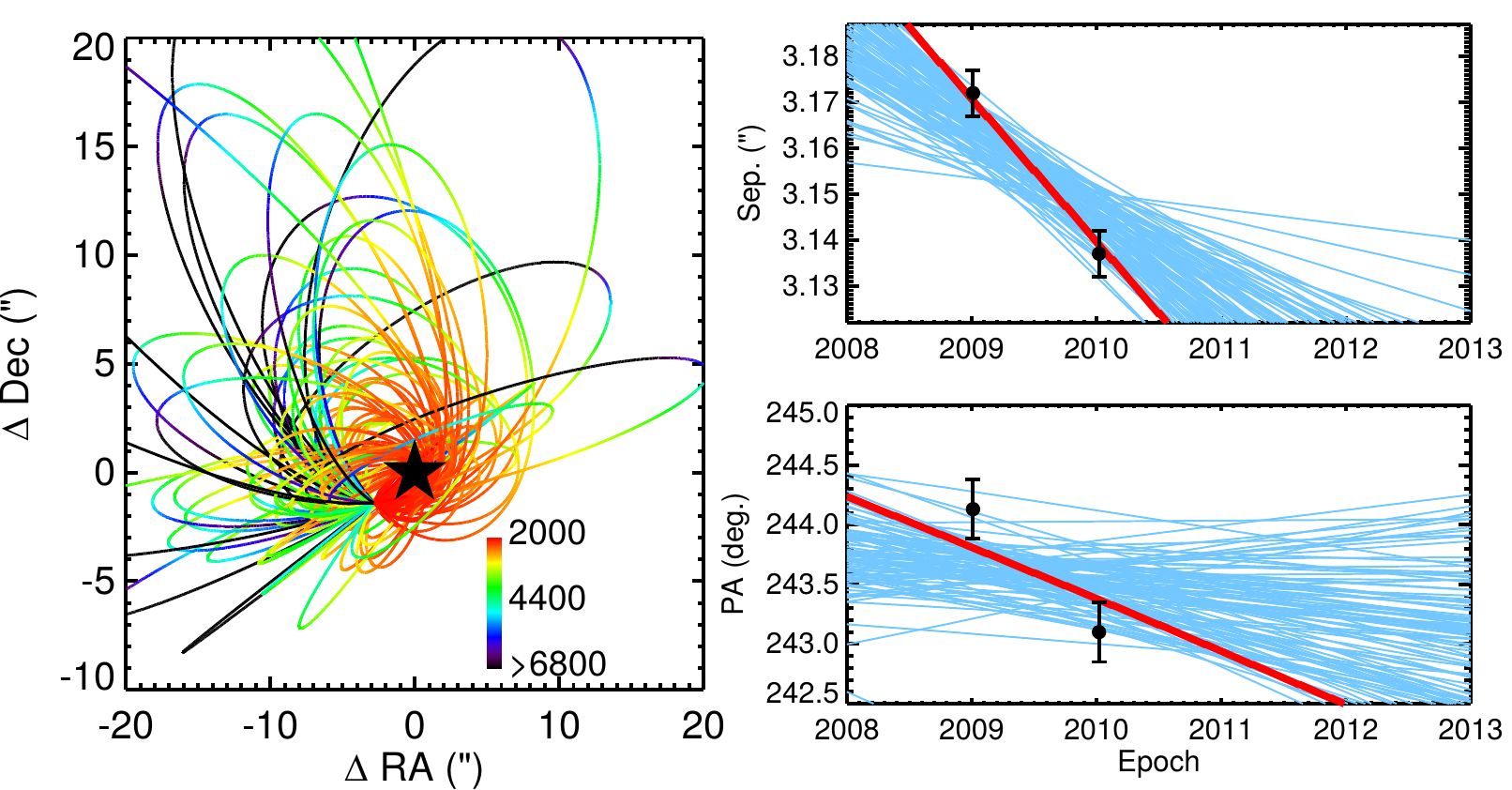}
    \caption{Depictions of 100 likely orbits fit to relative astrometry of CD-35~2722~B with respect to CD-35~2722~A. See figure \ref{fig:gj504_orbit}.}
    \label{fig:cd35_orbit}
\end{figure}

\end{document}